\documentclass[aps,pre,twocolumn,groupedaddress]{revtex4-1}
\usepackage{graphicx}
\usepackage{dcolumn}
\usepackage{bm}
\usepackage{amsmath}
\usepackage{amssymb}
\usepackage{color}

\begin{document}

\title{Binding of Curvature-Inducing Proteins onto Biomembranes}

\author{Hiroshi Noguchi}
\email[]{noguchi@issp.u-tokyo.ac.jp}
\affiliation{Institute for Solid State Physics, University of Tokyo, Kashiwa, Chiba 277-8581, Japan}

\begin{abstract}
We review the theoretical analyses and simulations 
of the interactions between curvature-inducing proteins and biomembranes.
Laterally isotropic proteins induce spherical budding,
whereas anisotropic proteins, such as  Bin/Amphiphysin/Rvs (BAR) superfamily proteins,
induce tubulation.
Both types of proteins can sense the membrane curvature.
We describe the theoretical analyses of various transitions of protein binding
accompanied by a change in various properties, such as the number of buds, the radius of membrane tubes, 
and the nematic order of anisotropic proteins.
Moreover, we explain the membrane-mediated interactions and protein assembly.
Many types of membrane shape transformations
(spontaneous tubulation, formation of polyhedral vesicles, polygonal tubes, periodic bumps, and network structures, etc.)
have been demonstrated by coarse-grained simulations.
Furthermore, traveling waves and Turing patterns under the coupling of reaction-diffusion dynamics and membrane deformation are described.
\end{abstract}


\maketitle

\section{Introduction}

In living cells, the biomembrane shapes of cells and organelles
are regulated by many types of proteins~\cite{mcma05,suet14,mcma15,joha15,mcma11,bran13,fain13,beth18,kaks18,schm11,anto16,raib09,hurl10,baum11,has21,stac13}.
Proteins are also involved in dynamic processes such as endocytosis, exocytosis, intracellular traffic, cell locomotion, and cell division.
Dysfunctions of these proteins have been implicated in neurodegenerative, cardiovascular, and neoplastic diseases.
Thus, it is important to understand the interactions of these proteins with biomembranes.

We review  theoretical and simulation studies on membrane shape transformations induced by protein binding
and self-assembly of the proteins.
We consider two types of curvature-inducing proteins: isotropic and anisotropic proteins.
Clathrin and coat protein complexes (COPI and COPII) bend membranes in a laterally isotropic manner and
 generate spherical buds~\cite{joha15,mcma11,bran13,fain13,beth18,kaks18},
and they can be modeled as laterally isotropic objects.
Polymer anchoring also induces a spontaneous curvature to gain the conformational entropy of polymer chains~\cite{bick06,hier96,auth03,auth05,evan03a,wern10}.
In contrast, Bin/Amphiphysin/Rvs (BAR) superfamily proteins bend the membrane anisotropically
and generate cylindrical membrane tubes~\cite{mcma05,suet14,joha15,itoh06,masu10,mim12a,fros08,guer09,sorr12,zhu12,tana13,daum14,adam15,tsuj21,snid21}.
Dynamins~\cite{schm11,anto16} and peptides of amphipathic $\alpha$-helices~\cite{drin10,gome16} also have anisotropic shapes and induce anisotropic curvatures.
These proteins can be modeled as anisotropic objects of a banana or crescent shape with a preferred bending orientation.

Studies using mean-field theory for the binding of isotropic proteins are reviewed in Sec.~\ref{sec:tiso}.
Protein binding models are described in Sec.~\ref{sec:sens}.
The budding of a vesicle and binding onto a tethered vesicle are described in Secs.~\ref{sec:bud} and \ref{sec:tet}, respectively.
Mean-field theory for the binding of anisotropic proteins is reviewed in Sec~\ref{sec:tani}.
The protein assemblies in membranes are reviewed in Sec.~\ref{sec:ass}.
Theories of 
Casimir-like interactions and the assembly of isotropic proteins are described
in Sec.~\ref{sec:asiso}.
The membrane-mediated interactions between two protein rods
and the assembly of anisotropic proteins are described
in Sec.~\ref{sec:asani}.
The membrane deformations caused by the mixture of multiple types of proteins
are described
in Sec.~\ref{sec:mix}.
In Sec.~\ref{sec:rd}, nonequilibrium membrane dynamics are described.
The membrane dynamics are coupled with a reaction-diffusion system of curvature-inducing proteins 
and exhibit Turing patterns and traveling waves.
Several types of coarse-grained membrane models have been developed;~\cite{muel06,vent06,nogu09}
here, we focus on recent simulation results of protein-induced membrane deformations using highly coarse-grained membrane models
that neglect the bilayer membrane structure:~\cite{nogu09}
the meshless membrane model~\cite{nogu06,shib11}
and dynamically triangulated membrane model~\cite{gomp04c,gomp97f,nogu05}.
The former model can apply to dynamics including topological change
and the latter is suitable to combine with continuum dynamics on the membrane surface such as a reaction-diffusion system.
Finally, an outlook is presented in Sec.~\ref{sec:out}.

\section{Theory of Isotropic Proteins}\label{sec:tiso}

\subsection{Local protein binding}\label{sec:sens}

\begin{figure}[tb]
\includegraphics[width=8.5cm]{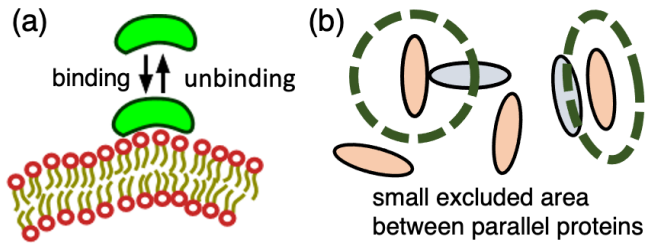}
\caption{
(Color online)
Schematics of protein binding.
(a) Binding and unbinding of a protein onto a membrane.
(b) An elliptic protein model for a mean-field theory.
A parallel protein pair has a smaller excluded area (represented by thick dashed lines)
than a  perpendicular pair, leading to a nematic order at a high density.
}
\label{fig:cat}
\end{figure}

First, we review mean-field theory for the binding of proteins with a laterally isotropic shape
 (i.e., no preferred bending direction).
The binding and unbinding of proteins are balanced at thermal equilibrium, as depicted in Fig.~\ref{fig:cat}(a).
The membrane bending rigidity and spontaneous curvature are modified by the binding,
and the bending free energy of a vesicle is given by~\cite{nogu21a}
\begin{eqnarray}\label{eq:Fcv0}
F_{\rm cv} =&&  4\pi\bar{\kappa}_{\rm d}(1-g_{\rm ves}) + \int {\rm d}A \ \Big\{ 2\kappa_{\rm d}H^2(1-\phi) \nonumber \\
 &+& \frac{\kappa_{\rm p}}{2}(2H-C_0)^2\phi  
+ (\bar{\kappa}_{\rm p}-\bar{\kappa}_{\rm d})K \phi  \Big\},
\end{eqnarray}
where $A$ is the membrane area,  $\phi$ is the local protein density ($\phi=1$ at the maximum coverage),
and $g_{\rm ves}$ is the genus of the vesicle.
$H$ and $K$ are the mean and Gaussian curvatures of each position, respectively,
i.e,  $H=(C_1+C_2)/2$ and $K=C_1C_2$, where
$C_1$ and $C_2$ are the principal curvatures.
The membrane is in a fluid phase, and
 $F_{\rm cv}$ is the second-order expansion to the curvature~\cite{canh70,helf73}.
The bare (protein-unbound) membrane has a bending rigidity of $\kappa_{\rm d}$ with zero  spontaneous curvature. 
The bound membrane has a bending rigidity of $\kappa_{\rm p}$ with finite spontaneous curvature $C_0$.
The first term of Eq.~(\ref{eq:Fcv0}) represents
the integral over the Gaussian curvature $K$
with the saddle-splay modulus $\bar{\kappa}_{\rm d}$  (also called the Gaussian modulus)~\cite{safr94}
of the bare membrane. This is determined by the membrane topology (Gauss--Bonnet theorem).
Lipid membranes typically have $\bar{\kappa}_{\rm d}/\kappa_{\rm d} \simeq -1$~\cite{hu12}.

\begin{figure*}[tb]
\includegraphics{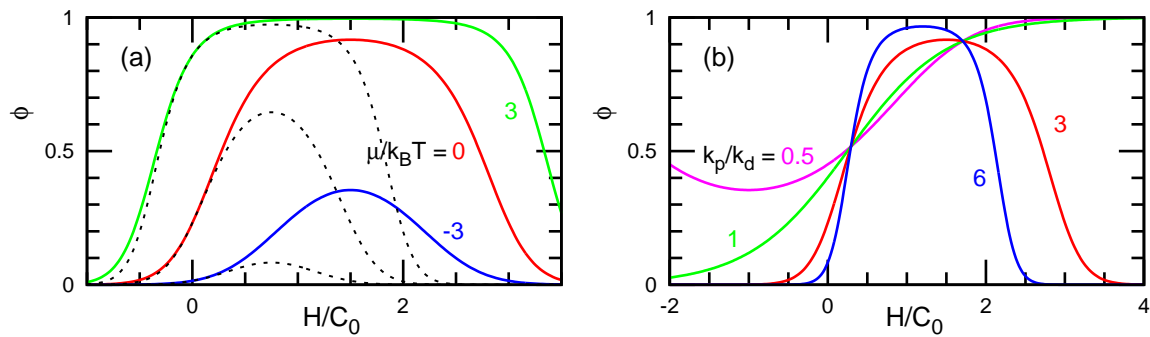}
\caption{
(Color online)
Protein density $\phi$ as a function of the local mean curvature $H$ 
at $C_0^2 a_{\rm p} = 0.04$ and  $\bar{\kappa}_{\rm p}/\kappa_{\rm p}=\bar{\kappa}_{\rm d}/\kappa_{\rm d}= -1$.
(a) The binding chemical potential $\mu$ is varied for  $\kappa_{\rm p}/\kappa_{\rm d}=3$. 
(b) The bending rigidity $\kappa_{\rm p}$ of the bound membrane is varied for  $\mu=0$.
The solid lines represent the data for spherical membranes ($H=C_1=C_2$)~\cite{nogu21a}.
The dashed lines in (a) represent the data for cylindrical membranes ($H=C_1/2$ and $C_2=0$).
}
\label{fig:phic}
\end{figure*}

In addition to the bending energy $F_{\rm cv}$, 
the free energy $F$ of a vesicle consists of 
the binding energy, inter-protein interaction energy, and mixing entropy:
\begin{eqnarray}\label{eq:F0}
F =&&  F_{\rm cv} + \int {\rm d}A \ \Big\{ - \frac{\mu}{a_{\rm p}}\phi   + b \phi^2   \nonumber \\
&+& \frac{k_{\rm B}T}{a_{\rm p}}[\phi \ln(\phi) +  (1-\phi) \ln(1-\phi) ] \Big\},
\end{eqnarray}
where 
 $a_{\rm p}$ is the area covered by one protein
(the maximum number of bound proteins is $A/a_{\rm p}$)
and  $k_{\rm B}T$ is the thermal energy.
The first term in the integral of Eq.~(\ref{eq:F0}) represents the protein binding energy,
and  $\mu$ is the chemical potential of the protein binding.
At higher $\mu$, more proteins bind to the membrane. 
The last two terms in Eq.~(\ref{eq:F0}) represent the pairwise inter-protein interactions and mixing entropy of the bound proteins, respectively.
The proteins have repulsive or attractive interactions at $b>0$ and $b<0$, respectively.
Note that $\chi\phi(1-\phi)$ is often used instead of $b\phi^2$, following Flory--Hagins theory~\cite{flor53,doi13} from polymer physics.
However, the parameter $\chi$ represents the interactions between different components (here, the bound and unbound membrane sites).
Since the interactions between proteins are typically dominant, $b\phi^2$ is more appropriate for protein binding.

In thermal equilibrium,
the protein density $\phi$ is locally determined for each membrane curvature.
When the inter-protein interactions are negligible ($b=0$),
 $\phi$ is expressed by a sigmoid function of $\mu$~\cite{nogu21a}:
\begin{eqnarray}\label{eq:phi0}
\phi &=& \frac{1}{1+\exp(w_{\rm b})}, \\ \nonumber
w_{\rm b}&=& - \frac{\mu}{k_{\rm B}T} + \frac{a_{\rm p}}{k_{\rm B}T}\big[2(\kappa_{\rm p}-\kappa_{\rm d})H^2 \\ \nonumber
&& + (\bar{\kappa}_{\rm p}-\bar{\kappa}_{\rm d})K  -2\kappa_{\rm p}C_0H  + \frac{\kappa_{\rm p}C_0^2}{2} \big].
\end{eqnarray}
When local binding and unbinding processes are considered as 
$d\phi/dt = \eta_{\rm b}(1-\phi) - \eta_{\rm ub}\phi$,
this corresponds to the relation between the binding and unbinding rates, 
$\eta_{\rm ub}/\eta_{\rm b}= \exp(w_{\rm b})$, i.e., the detailed balance at a local membrane region~\cite{gout21}.
For $b\ne 0$, $\phi$ is iteratively solved by replacing $w_{\rm b}$ with $w_{\rm b} + 2b\phi$ in Eq.~(\ref{eq:phi0})~\cite{nogu21a}.

Figure~\ref{fig:phic} shows that the protein binding depends 
on the local curvature $H$.
For a high curvature of $C_0$ or high rigidity of $\kappa_{\rm p}$,
the density $\phi$ changes steeply from $0$ to $1$ with a small increase in $H$. 
Here, $\phi$ exhibits a maximum value at the sensing curvature $H_{\rm s}$. 
For a spherical membrane, it is given by $H_{\rm s}= (\kappa_{\rm p}/2\kappa_{\rm dif})C_0$,
where $\kappa_{\rm dif}=\kappa_{\rm p}-\kappa_{\rm d}+(\bar{\kappa}_{\rm p}-\bar{\kappa}_{\rm d})/2$. 
The value of $H_{\rm s}$ is obtained from  $d\phi/dH =0$ using Eq.~(\ref{eq:phi0}) at $K=H^2$.
In contrast, 
the curvature generated by protein binding onto a free membrane is
$H_{\rm g}= [\kappa_{\rm p}\phi/2(\kappa_{\rm dif}\phi + \kappa_{\rm d})]C_0$, 
which is given by $dF/dH =0$.
Therefore, the  generated curvature is lower than the preferred curvature for the curvature sensing, since the membrane must bend together during the curvature generation.
For $\kappa_{\rm p}/\kappa_{\rm d}=\bar{\kappa}_{\rm p}/\bar{\kappa}_{\rm d}=1$,
 $\phi$ is a sigmoid function of $\mu$ so that more proteins bind to a higher curvature, even 
for $H \gg C_0$ (i.e., $H_{\rm s}=\infty$), as shown by the green line in Fig.~\ref{fig:phic}(b).
For a cylindrical membrane, the curvatures for curvature sensing and generation
are  $H_{\rm s}= [\kappa_{\rm p}/2(\kappa_{\rm p}-\kappa_{\rm d})]C_0$ 
and $H_{\rm g}=\{\kappa_{\rm p}\phi/2[(\kappa_{\rm p}-\kappa_{\rm d})\phi + \kappa_{\rm d}]\}C_0$, respectively.
These differences between the spherical and cylindrical membranes are 
due to the difference in the saddle-splay modulus, $\bar{\kappa}_{\rm p}-\bar{\kappa}_{\rm d}$
(compare the solid and dashed lines in Fig.~\ref{fig:phic}(a)).

Curvature-inducing proteins can sense the membrane curvature.
However, the sensing is not a sufficient condition for curvature generation.
Let us consider proteins or other molecules that reduce the bending rigidity (i.e., $\kappa_{\rm p}<\kappa_{\rm d}$) 
by remodeling the bound membrane, such as by reducing the membrane thickness.
They still sense the membrane curvature, 
but $\phi$ has a minimum instead of a maximum (see the magenta line in Fig.~\ref{fig:phic}(b)).
The bound membrane passively bends to allow other membrane parts to bend to their preferred curvatures,
since the total free energy is reduced.
Thus, the bound membrane may bend in the opposite direction to its preferred curvature.
Therefore, a larger bending rigidity ($\kappa_{\rm p}>\kappa_{\rm d}$) is significant for generating membrane curvature in a specific direction.

Finally, we explain two other bending-energy models
that are considered as subsets of the present model given by Eq.~(\ref{eq:Fcv0})~\cite{nogu21a}.
First, the protein bending energy is considered separately (often called curvature mismatch model).
The proteins adhere to the membrane surface,
and the membrane composition beneath the proteins remains almost unchanged.
In this case, the bending energy can be expressed as
\begin{eqnarray}\label{eq:Fcv1}
F_{\rm cv} =&&  4\pi\bar{\kappa}_{\rm d}(1-g_{\rm ves}) + \int {\rm d}A \ \Big\{ 2\kappa_{\rm d}H^2  \nonumber \\
&+&  [ \frac{\kappa_{\rm pa}}{2}(2H-C_{\rm 0a})^2 + \bar{\kappa}_{\rm pa}K ]\phi  \Big\}.
\end{eqnarray}
This bending energy is identical to Eq.~(\ref{eq:Fcv0}) with
  $\kappa_{\rm p}=\kappa_{\rm pa}+\kappa_{\rm d}$, $\bar{\kappa}_{\rm p}=\bar{\kappa}_{\rm pa}+\bar{\kappa}_{\rm d}$,
 $C_0= [\kappa_{\rm pa}/(\kappa_{\rm pa}+\kappa_{\rm d})] C_{\rm 0a}$, 
and the chemical potential shift $\Delta\mu = \mu - \mu_{\rm pa} = (\kappa_{\rm p}C_{\rm 0}^2-\kappa_{\rm pa}C_{\rm 0a}^2)a_{\rm p}/2$.
Here, $\kappa_{\rm pa}$ is the bending rigidity of the protein itself,
while $\kappa_{\rm p}$ is the rigidity, including the membrane beneath the protein.
This model can be used 
for  $\kappa_{\rm p}>\kappa_{\rm d}$, where the present model can be mapped to the curvature mismatch model.
This curvature mismatch model with $\bar{\kappa}_{\rm pa}=0$ was used in Refs.~\citenum{prev15,rosh17,frey20,tsai21}.

In some previous studies~\cite{sorr12,shi15,gov18,tozz19,kris19},
it was assumed that the bending rigidity is not modified by the binding
as a simple model, and
the following bending energy  was used:
\begin{equation}\label{eq:Fcv2}
F_{\rm cv} =   \int {\rm d}A \ \Big\{ \frac{\kappa_{\rm d}}{2}(2H-\phi C_{\rm 0})^2  \Big\}.
\end{equation}
This corresponds to the condition of $\kappa_{\rm p}=\kappa_{\rm d}$,  $\bar{\kappa}_{\rm p}=\bar{\kappa}_{\rm d}$, $b=\kappa_{\rm d}C_0^2/2$, 
and $\Delta\mu = \kappa_{\rm p}C_{\rm 0}^2a_{\rm p}/2$.
Here, neighboring proteins interact via the bending energy through $(\kappa_{\rm d}C_0^2/2)\phi^2$. 
Thus, this quadratic term is often neglected~\cite{rama00,shlo09}.
Since the linear and quadratic terms of $\phi$ represent membrane--protein and protein--protein interactions, respectively,
they should be treated differently.
For the same reason, 
one should not use the preaverages of both bending rigidity and spontaneous curvature 
as $F_{\rm cv} =  \int {\rm d}A\  (\kappa_{\rm d}+\kappa_1\phi)(2H-\phi C_{\rm 0})^2/2$ to represent membrane-protein interactions, 
since the higher-order terms can have dominant effects, as noted in Ref.~\citenum{nogu15b}.

\begin{figure*}[tb]
\includegraphics{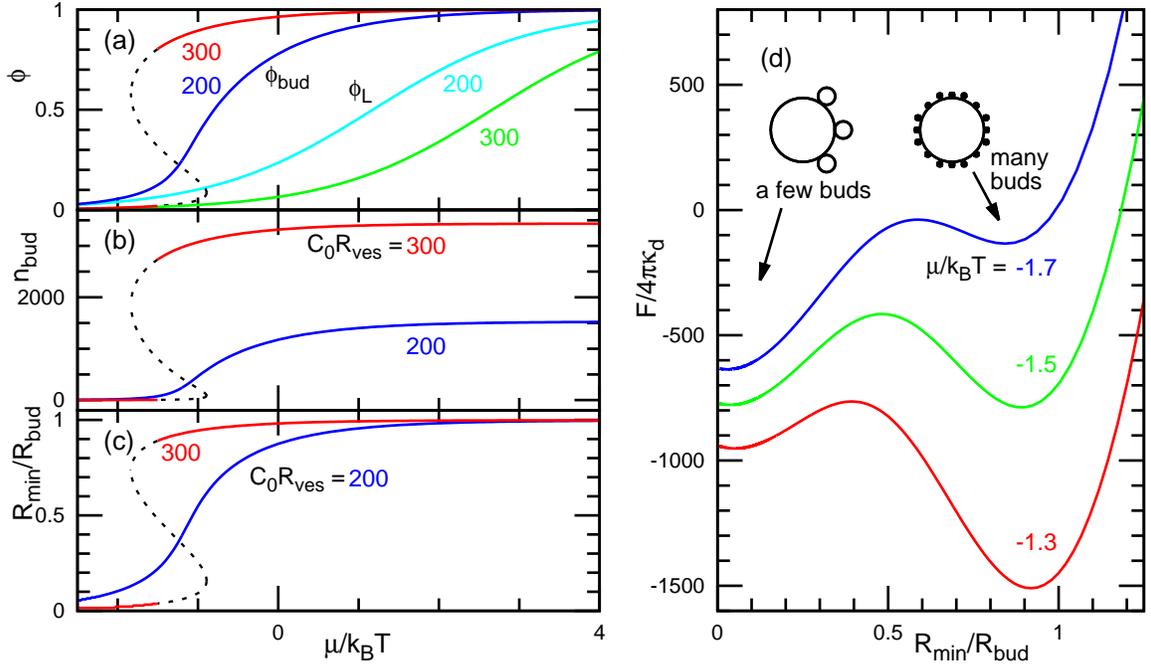}
\caption{
(Color online)
Budding induced by protein binding
at $V^*=0.9$, $\kappa_{\rm p}/\kappa_{\rm d}= 3$, $\bar{\kappa}_{\rm p}/\kappa_{\rm p}=\bar{\kappa}_{\rm d}/\kappa_{\rm d}=  -1$,
and $a_{\rm p}/R_{\rm ves}^2 = 10^{-6}$. 
(a)--(c) Binding chemical potential $\mu$ dependence of (a) protein density $\phi$, 
(b) the number $n_{\rm bud}$ of buds, and (c) the bud curvature $1/R_{\rm bud}$
for  $C_0R_{\rm ves}= 200$ and $300$. 
$\phi_{\rm bud}$ and $\phi_{\rm L}$ are the protein densities of the buds and main spherical component
of the vesicle, respectively.
$R_{\rm bud}$ is the radius of the buds
and  $R_{\rm min}= 2(\kappa_{\rm dif}+\kappa_{\rm d})/\kappa_{\rm p}C_0$
 is the radius of the minimum bud curvature energy at $\phi_{\rm bud}=1$.
(d) Free-energy profile for the bud curvature $1/R_{\rm bud}$ 
for  $\mu/k_{\rm B}T=-1.7$, $-1.5$, and $-1.3$ at  $C_0R_{\rm ves}= 300$. 
The vesicle exhibits a first-order transition between a few large buds and many small buds.
Adapted from Ref.~\citenum{nogu21a}.
}
\label{fig:bud}
\end{figure*}

\subsection{Budding}\label{sec:bud}

In endo/exocytosis and intracellular traffic,
protein binding induces membrane budding, leading to the formation of a spherical vesicle.
Clathrins assemble on a membrane and form a spherical bud with  $20$--$200$-nm diameter~\cite{mcma11,kaks18,avin15,sale15}.
COPI and COPII coat vesicles for retrograde and anterograde transports between the Golgi apparatus and endoplasmic reticulum (ER), respectively,
and form spherical bud with $60$--$100$-nm diameters, under typical conditions~\cite{bran13,fain13,beth18}.
This budding can be understood by a protein model with an isotropic spontaneous  curvature.
The budding process has been theoretically analyzed using a spherical-cap geometry~\cite{frey20,lipo92,sens03} 
and more detailed geometry~\cite{fore14}.

Since bud radii are much smaller than the size of liposomes and cells in typical experiments,
many buds are formed.
Here, we consider bud formation in a $\mu$m-size vesicle~\cite{nogu21a}.
The budded vesicle is modeled as connected spheres, as depicted in the inset of Fig.~\ref{fig:bud}(d).
The buds have the same radius $R_{\rm bud}$, and the number of the buds is $N_{\rm bud}$.
For a vesicle with a constant surface area $A$ and volume $V$,
one degree of freedom is left, since three variables exist: $R_{\rm bud}$, $N_{\rm bud}$, and the radius of the large spherical component. 
Hence, the free energy minimum can be easily solved by Eq.~(\ref{eq:F0}).
Figure~\ref{fig:bud} shows the budding transition at the reduced volume $V^*=V/(4\pi R_{\rm ves}^3/3)= 0.9$,
where $R_{\rm ves}= \sqrt{A/4\pi}$.
More buds with a smaller radius are formed
with increasing protein density $\phi_{\rm bud}$ of the buds, as the binding chemical potential $\mu$ increases.
For a large spontaneous curvature ($C_0R_{\rm ves}=300$), a first-order transition occurs between a few buds with a large radius
and many buds with a small radius (see red curves in Figs.~\ref{fig:bud}(a)--(c) and Fig.~\ref{fig:bud}(d)),
whereas continuous changes are obtained for a small spontaneous curvature ($C_0R_{\rm ves}=200$).
In addition, the effects of inter-protein interactions, protein-insertion-induced area expansion,
 variation in the saddle-splay-modulus, and the area-difference-elasticity energy~\cite{seif97,svet09,svet89} 
were investigated in Ref.~\citenum{nogu21a}.
This scheme, using a simple geometry, is easily applicable to other shape transformations, as described in the next subsection
for a tethered vesicle.

\subsection{Binding onto tethered vesicle}\label{sec:tet}

To examine the curvature sensing of protein binding,
a tethered vesicle has been most widely employed~\cite{baum11,has21,sorr12,prev15,rosh17,tsai21,dimo14,more19,roux10,aimo14}.
Using optical tweezers and a micropipette,
a narrow membrane tube (tether) is extended from a spherical vesicle.
The tube radius can be varied by the mechanical force imposed by the optical tweezers.
BAR proteins~\cite{baum11,sorr12,prev15,tsai21}, G-protein coupled receptors~\cite{rosh17}, annexins~\cite{more19}, a dynamin~\cite{roux10}, ion channel~\cite{aimo14}, and Ras protein~\cite{lars20} 
have been reported to exhibit curvature sensing.
Alternatively, the curvature sensing can be examined to measure the binding probability to different sizes of spherical vesicles~\cite{lars20,hatz09,zeno19}.
The difference in the sensing between the tubes and spherical vesicles was reported in Ref.~\citenum{lars20}.
For isotropic proteins, this difference can be caused by the Gaussian curvature term (see Eq.~(\ref{eq:phi0}) and Fig.~\ref{fig:phic}(a)).

\begin{figure*}[tb]
\includegraphics{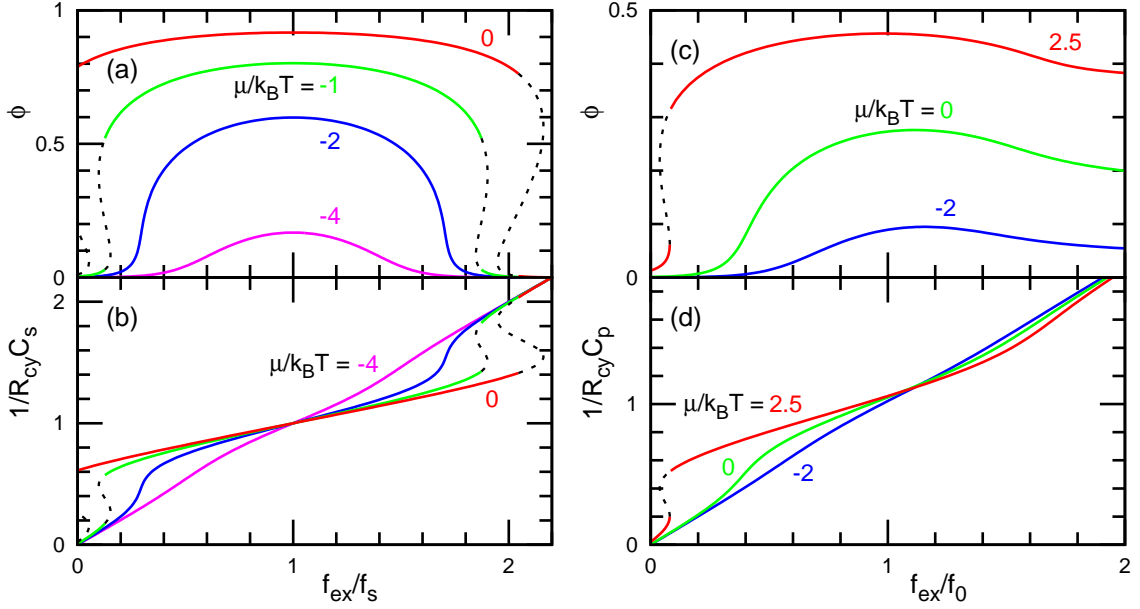}
\caption{
(Color online)
Protein binding onto a membrane tube.
Protein density $\phi$ and tube curvature $1/R_{\rm cy}$ dependence on the axial force $f_{\rm ex}$
are shown in (a),(c) and in (b),(d), respectively.
(a),(b) Isotropic proteins for $\mu/k_{\rm B}T=-4$, $-2$, $-1$, and $0$ at $\kappa_{\rm p}/\kappa_{\rm d}= 3$
and $a_{\rm p}C_0^2 = 0.16$. 
(c),(d) Anisotropic proteins for $\mu/k_{\rm B}T=-2$, $0$, and $2.5$
  at $\kappa_{\rm pm}/k_{\rm B}T=60$, $\kappa_{\rm pside}=0$, $d_{\rm el}=3$, and  $a_{\rm p}{C_{\rm p}}^2 = 0.26$. 
The solid lines represent equilibrium states.
The black dashed lines represent metastable and free-energy barrier states.
The isotropic proteins exhibit a first-order transition twice at large $\mu$~\cite{nogu21b}.
 In contrast, the anisotropic proteins exhibit it only once~\cite{nogu22}.
}
\label{fig:tube}
\end{figure*}

A tethered vesicle can be modeled by a spherical membrane connected to a cylindrical tube~\cite{smit04,nogu21b}.
Under the constraints of the total area $A$ and volume $V$,
the shape change is expressed by one variable, as in the vesicle budding in Sec.~\ref{sec:bud}.
Moreover, the volume change in the tube  is neglected for a narrow tube with $\pi {R_{\rm cy}}^2L_{\rm cy} \ll V$,
where  $R_{\rm cy}$ and $L_{\rm cy}$ are the radius and length of the tube, respectively.
In this limit, the axial force $f_{\rm ex}$ to extend the tube is expressed as~\cite{nogu21b}
\begin{eqnarray}
f_{\rm ex} &=& \frac{\pi[(\kappa_{\rm p}-\kappa_{\rm d})\phi+\kappa_{\rm d}]L_{\rm cy}}{(1 - (V^*)^{2/3}){R_{\rm ves}}^2}
- 2\pi\kappa_{\rm p}C_0\phi  \label{eq:fext1} \\ \label{eq:fext2}
 &=& \frac{2\pi[(\kappa_{\rm p}-\kappa_{\rm d})\phi+\kappa_{\rm d}]}{R_{\rm cy}}
- 2\pi\kappa_{\rm p}C_0\phi  \\ \label{eq:fext3}
 &=&  \Big\{ \Big[ (\frac{\kappa_{\rm p}}{\kappa_{\rm d}}-1)\phi+1 \Big]\Big(\frac{1}{R_{\rm cy}C_{\rm s}}  - 1\Big) +1 \Big\} f_{\rm s},
\end{eqnarray}
where the sensing tube curvature $C_{\rm s} =  \kappa_{\rm p}C_0/(\kappa_{\rm p}-\kappa_{\rm d})$  and
 corresponding force $f_{\rm s} = 2\pi \kappa_{\rm d}C_{\rm s}$.
The protein density $\phi$ on the membrane tube is determined by Eq.~(\ref{eq:phi0}) with $C_1=1/R_{\rm cy}$ and $C_2=0$.
The force dependence curves of $\phi$ are reflection symmetric with respect to $f_{\rm ex}/f_{\rm s}=1$ and take maxima at $f_{\rm ex}/f_{\rm s}=1$ (see Fig.~\ref{fig:tube}(a)). 
The tube curvature $1/R_{\rm cy}$ is point symmetric with respect to $f_{\rm ex}/f_{\rm s}=1$ (see Fig.~\ref{fig:tube}(b)).
For completely unbound tubes ($\phi=0$),
$f_{\rm ex}$ is proportional to $1/R_{\rm cy}$  as $f_{\rm ex}/f_{\rm s}  = 1/R_{\rm cy}C_{\rm s}$.
For large  $\mu$, first-order transitions occur twice between the unbound and bound tubes with different tube radii.
These two transitions appear symmetrically with respect to $f_{\rm ex}/f_{\rm s}=1$.

The surface tension  $\gamma$ is obtained from $\partial G/\partial R_{\rm cy}|_{L_{\rm cy}}=0$,
where $G=F + \gamma A$~\cite{nogu21b}:
\begin{eqnarray} \label{eq:tent}
\gamma =&& \frac{(\kappa_{\rm p}-\kappa_{\rm d})\phi+\kappa_{\rm d}}{2{R_{\rm cy}}^2} 
+  \Big(\frac{\mu}{a_{\rm p}} - \frac{\kappa_{\rm p}}{2}{C_0}^2\Big)\phi - b \phi^2 \nonumber \\
 &-&  \frac{k_{\rm B}T}{a_{\rm p}}[\phi \ln(\phi) +  (1-\phi) \ln(1-\phi) ]. 
\end{eqnarray}
In this narrow-tube limit, the Laplace tension is neglected~\cite{nogu21b}.
For completely unbound membrane tubes of $\phi=0$, a well-known relation
$\gamma=\kappa_{\rm d}/2{R_{\rm cy}}^2 = f_{\rm ex}/4\pi R_{\rm cy}$  is obtained.
This relation has been used to experimentally estimate $R_{\rm cy}$ and $\kappa_{\rm d}$ from $\gamma$ and $f_{\rm ex}$~\cite{dimo14,bo89,evan96,cuve05}.
Protein binding largely modifies the surface tension.

This theory reproduces the meshless simulation results for a homogeneous phase very well,
 including the first-order transition~\cite{nogu21b}.
In addition, membrane deformation can induce microphase separation near the transition condition, as described in Sec.~\ref{sec:isoves}.

\begin{figure*}[tb]
\includegraphics[width=16cm]{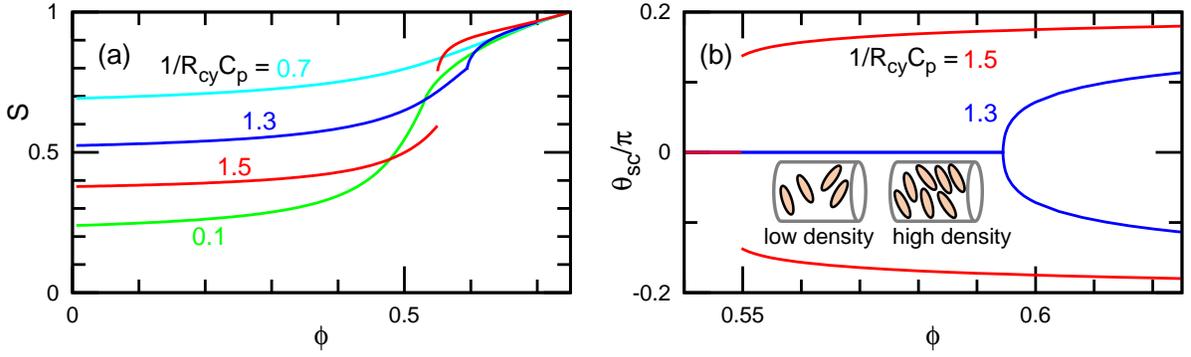}
\caption{
(Color online)
Equilibrium states of the anisotropic proteins on a membrane tube  at $\kappa_{\rm pm}/k_{\rm B}T=40$, $\kappa_{\rm pside}=0$, 
 $d_{\rm el}=3$, and  $a_{\rm p}{C_{\rm p}}^2 = 0.26$~\cite{nogu22}.
(a) Orientational degree $S$ for $1/R_{\rm cy}C_{\rm p}=0.1$, $0.7$, $1.3$, and $1.5$.
(b) Angle $\theta_{\rm sc}$ between the nematic order and azimuthal direction for $1/R_{\rm cy}C_{\rm p}=1.3$ and $1.5$.
Second-order and first-order transitions occur for $1/R_{\rm cy}C_{\rm p}=1.3$ and $1.5$, respectively.
The protein orientations are schematically depicted for low and high protein densities $\phi$ in (b).
}
\label{fig:ang}
\end{figure*}

\begin{figure*}[tb]
\includegraphics[width=16cm]{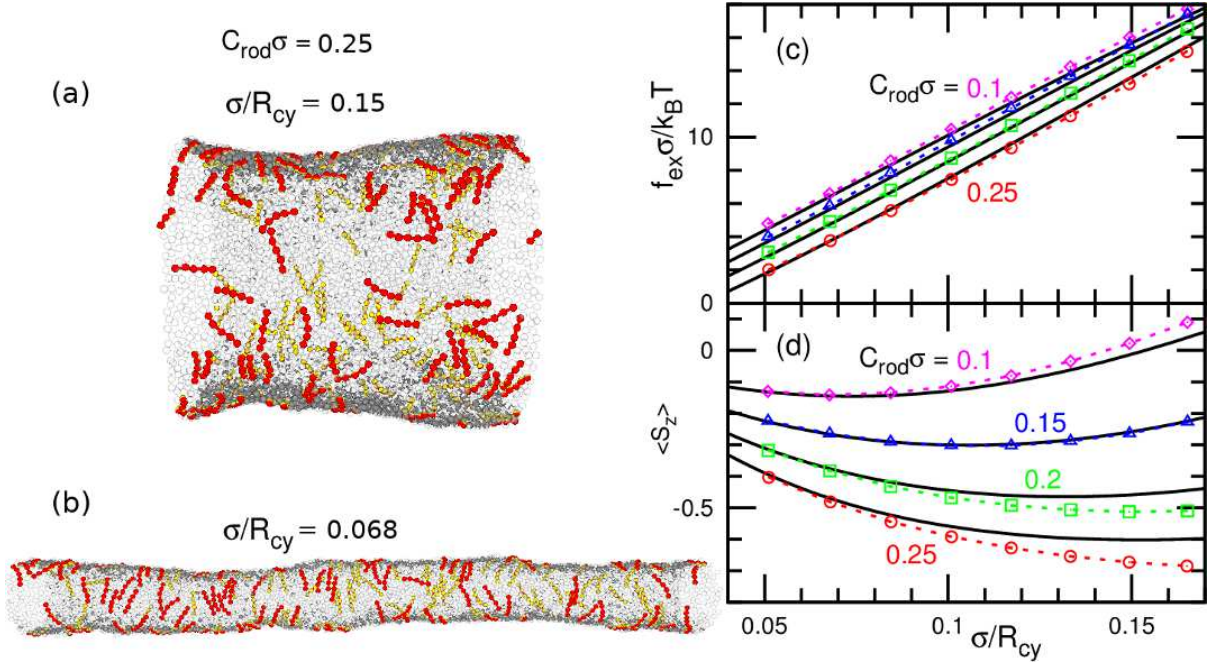}
\caption{
(Color online)
Membrane tubes with protein rods of length $r_{\rm rod} = 5\sigma$ 
at $\phi=0.167$.
(a),(b) Snapshots of a meshless membrane simulation for (a) $r_{\rm rod}/R_{\rm cy}= 0.75$ and 
(b)  $r_{\rm rod}/R_{\rm cy}= 0.34$ at $C_{\rm rod}r_{\rm rod} = 1.25$.
A protein rod is displayed as
a chain of spheres whose halves are colored
in red and yellow.
The orientation of a protein segment lies along the direction from the 
yellow to red hemispheres.
Transparent gray particles represent
membrane particles.
(c),(d) Dependence of (c) the axial force $f_{\rm ex}$ and (d) orientational order $\langle S_z\rangle$ along the membrane tube 
on the tube radius $R_{\rm cy}$ for $C_{\rm rod}r_{\rm rod} = 0.5$, $0.75$, $1$, and $1.25$.
The symbols with dashed lines represent the simulation data.
The black solid lines represent the theoretical results.
(c),(d) Reproduced from Ref.~\citenum{nogu22} with permission from the Royal Society of Chemistry.
}
\label{fig:n80}
\end{figure*}

\section{Theory of Anisotropic Proteins}\label{sec:tani}

In this section, we describe a mean-field theory for the binding of anisotropic proteins~\cite{tozz21,roux21,nogu22}.
The bound protein is assumed to have an elliptical shape laterally on the membrane,
and the orientation-dependent excluded area is considered based on Nascimentos' theory for three-dimensional liquid-crystals~\cite{nasc17}.
The proteins can be aligned by the inter-protein interactions and preferred bending direction.
The degree of orientational order $S$ is given by
$S = 2 \langle s_{\rm p}(\theta_{\rm ps}) \rangle$, where
$s_{\rm p}(\theta_{\rm ps}) = \cos^2(\theta_{\rm ps}) - 1/2$,
 $\langle ... \rangle$ is the ensemble average,
and $\theta_{\rm ps}$ is the angle between the major protein axis and nematic orientation {\bf S}.

The protein has an anisotropic bending energy described by
\begin{eqnarray} \label{eq:ubrod}
U_{\rm p} &=&   \frac{\kappa_{\rm pm}a_{\rm p}}{2}(C_{\ell 1} - C_{\rm p})^2 + \frac{\kappa_{\rm pside}a_{\rm p}}{2}(C_{\ell 2} - C_{\rm pside})^2, \nonumber \\
C_{\ell 1} &=& C_1 \cos^2( \theta_{\rm pc} ) + C_2  \sin^2(\theta_{\rm pc} ),  \\
C_{\ell 2} &=& C_1 \sin^2( \theta_{\rm pc} ) + C_2  \cos^2( \theta_{\rm pc} ), \nonumber
\end{eqnarray}
where $\kappa_{\rm pm}$ and $C_{\rm p}$ are the bending rigidity and spontaneous curvature along the major protein axis, respectively,
and  $\kappa_{\rm pside}$ and $C_{\rm pside}$ are those along the minor axis (side direction).
$\theta_{\rm pc}=\theta_{\rm ps}-\theta_{\rm sc}$ is the angle between the protein major axis and the eigenvector of the principal curvature $C_1$ of the membrane,
where $\theta_{\rm sc}$ is the angle between {\bf S} and  the eigenvector.
The protein area is $a_{\rm p} = \pi \ell_1\ell_2/4$,
where $\ell_1$ are $\ell_2$ are the lengths in the major and minor axes, respectively.

The free energy $F_{\rm p}$ of the bound proteins is expressed as~\cite{tozz21}
\begin{eqnarray}
F_{\rm p} &=&  \int {\rm d}A\ \frac{\phi k_{\rm B}T}{a_{\rm p}}\Big[\ln(\phi) + \frac{S \Psi}{2} \nonumber \\
&&- \ln\Big(\int_{-\pi}^{\pi} w(\theta_{\rm ps})\ {\rm d}\theta_{\rm ps}\Big)\Big], 
\end{eqnarray}
where
\begin{eqnarray}
w(\theta_{\rm ps})  &=&  g\exp[\Psi s_{\rm p}(\theta_{\rm ps}) + \bar{\Psi}\sin(\theta_{\rm ps})\cos(\theta_{\rm ps})  \nonumber \\
&&- U_{\rm p}/k_{\rm B}T ]\Theta(g), \\
g   &=& 1-\phi (b_0-b_2S s_{\rm p}(\theta_{\rm ps})),
\end{eqnarray}
$\Psi$ and $\bar{\Psi}$ are the symmetric  and asymmetric components of the nematic tensor, respectively, 
and  $\Theta(x)$ denotes the unit step function.
The factor $g$ expresses the effect of the orientation-dependent excluded volume interaction.
When two proteins are oriented parallel to each other,
the excluded area $A_{\rm exc}$ between them is smaller than that of perpendicular pairs, as shown in Fig.~\ref{fig:cat}(b).
This difference increases with increasing aspect ratio $d_{\rm el}=\ell_1/\ell_2$.
The area  $A_{\rm exc}$ is approximated as $A_{\rm exc}= (b_0 - b_2(\cos^2(\theta_{\rm pp})-1/2))a_{\rm p}/\lambda$,
where $\theta_{\rm pp}$ is the angle between the major axes of two proteins.
The factor $\lambda$ represents the packing ratio,
and the maximum density is given by  $\phi_{\rm max}=1/\lambda(b_0-b_2/2)$. 

\begin{figure*}[tb]
\includegraphics[width=16cm]{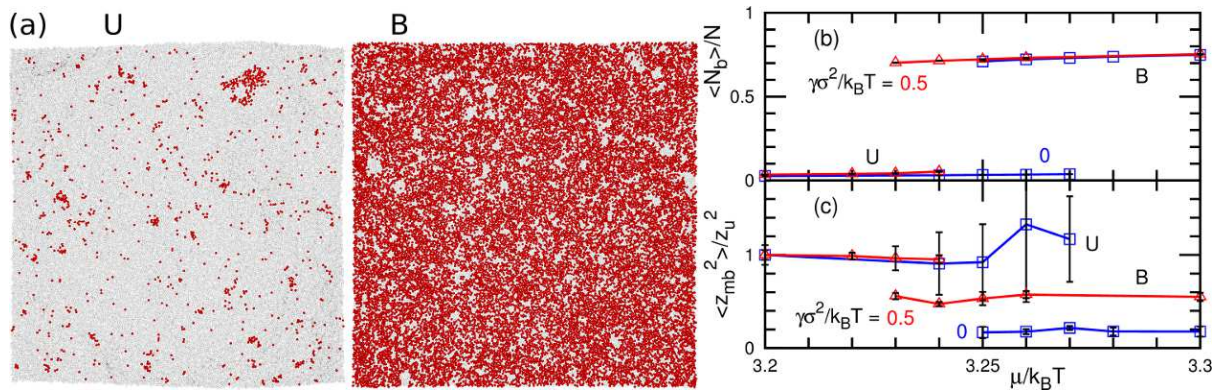}
\caption{
(Color online)
Binding transition of isotropic proteins with $\kappa_{\rm p}/\kappa_{\rm d}\simeq 9$ and $C_0=0$.
(a) Snapshots of the meshless membrane simulation at $\mu/k_{\rm B}T=3.26$ and $\gamma=0$.
Red and transparent spheres represent bound and unbound membranes, respectively.
The unbound (U) and bound (B) phases coexist.
(b),(c) Chemical potential $\mu$ dependence of (b) the mean ratio of bound membranes $\langle N_{\rm b}\rangle/N$  
and (c) the mean membrane vertical spans $\langle {z_{\rm mb}}^2\rangle$ at $\gamma\sigma^2/k_{\rm B}T=0$ and $0.5$,
where  ${z_{\rm u}}^2$ are the values of $\langle {z_{\rm mb}}^2\rangle$ for the U phase at $\mu/k_{\rm B}T=3.2$.
The first-order transition occurs between the U and B phases.
Adapted from Ref.~\citenum{gout21} with permission from the Royal Society of Chemistry.
}
\label{fig:casi}
\end{figure*}

For a flat membrane, the proteins exhibit an isotropic-to-nematic transition~\cite{tozz21}.
For a membrane tube with a curvature of $0.1C_{\rm p}\lesssim 1/R_{\rm cy}\lesssim C_{\rm p}$,
the orientational order $S$ continuously increases  with increasing $\phi$ 
(see Fig.~\ref{fig:ang}(a)).
For narrow tubes with $1/R_{\rm cy}> C_{\rm p}$,
the preferred protein orientation is tilted from the azimuthal direction.
At low $\phi$, proteins tilted in the left and right directions coexist equally such that $\theta_{\rm sc}=0$ or $\pi/2$.
Meanwhile, at high $\phi$, only one type of tilt can exist owing to 
the orientation-dependent excluded volume interaction.
Thus, a transition occurs between these two states.
At $1/R_{\rm cy}C_{\rm p}=1.3$ and $1.5$, this transition is the second-order and first-order (see Fig.~\ref{fig:ang}(b)),
and the slope and position of $S$ exhibit corresponding gaps, respectively~\cite{nogu22} (see Fig.~\ref{fig:ang}(a)).

The equilibrium protein density on the membrane tube is obtained by minimizing $F=F_{\rm p} + F_{\rm mb} - f_{\rm ex}L_{\rm cy} - \mu N_{\rm p}$~\cite{nogu22}.
The $f_{\rm ex}$ dependence curves of  $\phi$ and $1/R_{\rm cy}$ are not symmetric, unlike for isotropic proteins
(compare Figs.~\ref{fig:tube}(c),(d) with \ref{fig:tube}(a),(b)).
The first-order transition between narrow and wide tubes with different densities occurs only once.
The small variation at $f_{\rm ex}>f_0$ is due to the tilt of the proteins in narrow tubes,
where $f_0= 2\pi \kappa_{\rm d}C_{\rm p}$ is the force at $1/R_{\rm cy}=C_{\rm p}$.
Interestingly, significant effects are observed even at $f_{\rm ex}<f_0$.
The sensing curvature giving the maximum of $\phi$ decreases with increasing $\mu$.
This dependence of the sensing curvature has been observed in experiments on BAR superfamily proteins~\cite{prev15,tsai21}.
The tube curvature giving the maximum of $S$ is different from the sensing curvature
and varies from $C_{\rm p}$ to $C_{\rm p}/2$ with increasing $\mu$ and/or $\kappa_{\rm pm}$.

This theory reproduces the simulation results of short protein rods well, as shown in Fig.~\ref{fig:n80}~\cite{nogu22}.
The protein consists of five particles and the protein contour length is $r_{\rm rod}=5\sigma$.
Hereinafter, we consider a particle diameter of $\sigma$.
Here, the orientational order $S_z$ along the tube axis is shown, since it is more easily measured in simulations and experiments.
For $\theta_{\rm sc}=0$ and $\pi/2$, $S_z=-S$ and $S_z=S$, respectively.
Since the protein rods are flexible in these simulations, the effective curvature $C_{\rm p}$ is slightly smaller than the rod curvature $C_{\rm rod}$.
For long protein rods with  $r_{\rm rod}=10\sigma$, theory and simulation results show less agreement
owing to cluster formation by membrane-mediated attraction between the proteins.

\section{Protein Assembly}\label{sec:ass}

The mean-field theories described in the previous two sections
assume a uniform distribution of proteins in each membrane component.
However, proteins often form assemblies that change the membrane shape.
In this section, we review protein assembly.

\subsection{Isotropic proteins}\label{sec:asiso}

\subsubsection{Casimir-like interactions}\label{sec:casi}

First, we consider isotropic proteins with zero spontaneous curvature (Eq.~(\ref{eq:Fcv0}) with $C_0=0$).
For a flat membrane, the bending energy does not play any role 
in the mean-field theory described in Sec.~\ref{sec:tiso}.
However, membrane fluctuations depend on the bending rigidity.
For an unbound membrane ($\phi=0$), the membrane height fluctuates as $\langle |h_q|^2 \rangle= k_{\rm B}T/(\gamma q^2 + \kappa_{\rm d} q^4)$ in thermal equilibrium,
where $h_q$ is the Fourier transform of the membrane height in the Monge representation~\cite{safr94}.
The surface tension calculated by this spectrum corresponds to the mechanical frame tension $\gamma$ conjugated to the projected area in the range of usual statistical errors~\cite{shib16}.
Note that the internal tension conjugated to the real membrane area is slightly larger than the mechanical tension
owing to the excess membrane area from the projected area~\cite{shib16,davi91,fara03a,gueg17}.

\begin{figure*}[t]
\includegraphics[width=17cm]{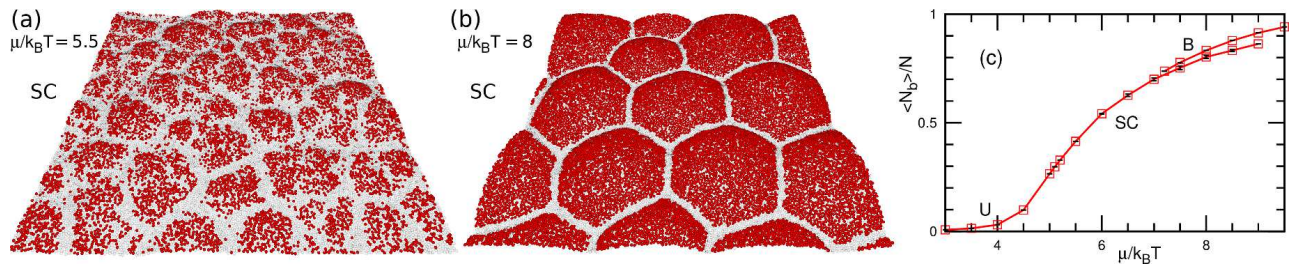}
\caption{
(Color online)
Binding of isotropic proteins with a finite spontaneous curvature $C_0\sigma=0.1$  onto a flat membrane at $\kappa_{\rm p}/\kappa_{\rm d}\simeq 9$ and $\gamma\sigma^2/k_{\rm B}T=0.5$.
(a),(b) Snapshots of the separated/corrugated (SC) phase at (a) $\mu/k_{\rm B}T=5.5$ and (b) $\mu/k_{\rm B}T=8$.
Red and transparent gray spheres represent bound and unbound membranes, respectively.
(c) Chemical potential $\mu$ dependence of (b) the mean ratio of bound membranes $\langle N_{\rm b}\rangle/N$.
The second- and first-order transitions occur between the U and SC phases and between the SC and B phases, respectively.
Adapted from Ref.~\citenum{gout21} with permission from the Royal Society of Chemistry.
}
\label{fig:psep}
\end{figure*}

Membrane fluctuations are reduced at high $\gamma$ and/or high $\kappa_{\rm d}$, i.e., the conformational entropy of the membrane decreases.
Since rigid proteins of high $\kappa_{\rm p}$ reduce the fluctuations of surrounding unbound membranes,
the proteins have attractive interactions with each other to increase the membrane conformational entropy~\cite{gout21,goul93}.
This Casimir-like interaction is expressed by the sum of $-6k_{\rm B}T(a/r_{ij})^4$ in leading order,
where the $r_{ij}$ are the distances between two rigid proteins, and $a$ is the protein size.
Thus, bound and unbound membranes are effectively repelled.
As a result, the protein binding exhibits a first-order transition between bound and unbound membranes,
as shown in Fig.~\ref{fig:casi}~\cite{gout21}.
The membrane vertical span ${z_{\rm mb}}^2=\sum_{i}^{N} (z_i-z_{\rm G})^2/N$
is significantly changed by the transition,
where $z_{\rm G}=\sum_{i}^{N} z_i/N$. 
At higher tension $\gamma$, the coexistence range of the chemical potential $\mu$ is reduced.

In the present case, the Casimir-like interaction is  well approximated as a pairwise additive. 
In contrast, the Casimir-like interaction between anchoring molecules connecting neighboring membranes
 is not well approximated by pairwise interactions owing to screening effects~\cite{weil10,nogu13}.

For transmembrane proteins, such as G-protein-coupled receptors and channel proteins,
the lengths of the protein hydrophobic domains can differ from the thickness of the surrounding membrane~\cite{lee03,ande07,vent05}.
To adjust their heights, the lipid membrane deforms, or alternatively, the domains are tilted.
This hydrophobic mismatch generates an additional interaction between proteins~\cite{phil09,deme08,schm08,four99}.

\begin{figure*}[tb]
\includegraphics[width=16cm]{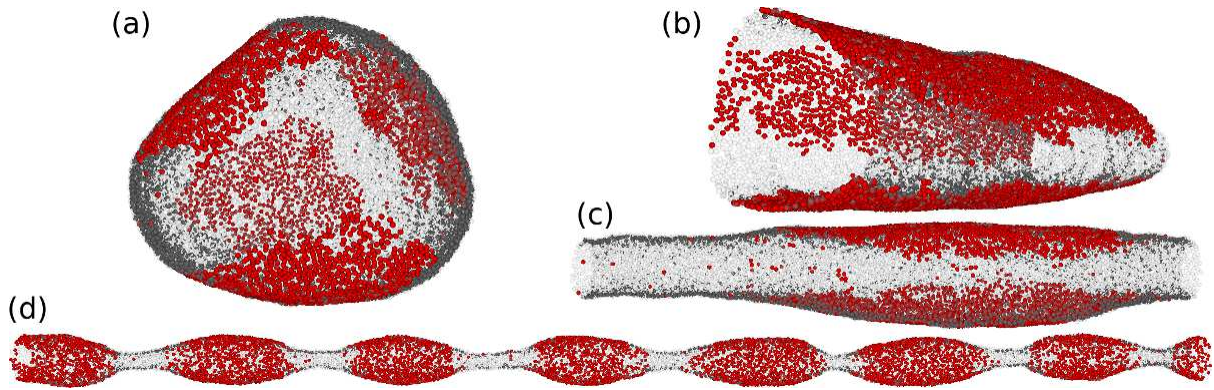}
\caption{
(Color online)
Snapshots of a vesicle and tubular membranes bound by isotropic proteins~\cite{nogu16a,nogu21b}.
for (a)--(c) $\kappa_{\rm p}/\kappa_{\rm d}\simeq 9$ and $C_0\sigma=0$ and for (d) $\kappa_{\rm p}/\kappa_{\rm d}\simeq 4$ and $C_0\sigma=0.05$.
(a) Tetrahedral vesicle at binding chemical potential $\mu/k_{\rm B}T=3.4$. 
(b) Triangular tube at $\mu/k_{\rm B}T=3.5$ and $L_{z}/\sigma=100$.
(c) Disk-shaped tube connecting with unbound circular tube at $\mu/k_{\rm B}T=3.5$ and  $L_{z}/\sigma=150$.
(d) Beaded-necklace-like tube at $\mu/k_{\rm B}T=3$ and $f_{\rm ex}\sigma/k_{\rm B}T=38$.
The disk-shaped tubes are also stable under the conditions of (b).
The tube length is fixed in (b) and (c), while the axial force is applied in (d).
Red and transparent gray spheres represent bound and unbound membranes, respectively.
}
\label{fig:iso}
\end{figure*}

\subsubsection{Domain formation by bending deformation}\label{sec:isoves}

When proteins and other membrane inclusions induce local bending of the bound membrane,
the situation becomes more complicated.
The interactions between spherical colloids bound on a membrane
have been intensively studied~\cite{dasg17,sari13,reyn07,auth09,sari12,barb18}.
They have a repulsive interaction, which is weakened by positive surface tension.
A membrane bound by colloids bends with a fixed curvature.
In contrast, proteins are soft objects, and the curvature of the bound membrane 
can largely deviate from the protein-preferred curvature.
This finite bending rigidity $\kappa_{\rm p}$ of the bound membrane is an important factor for protein binding
and the resulting membrane deformation.

In a flat membrane, the binding of proteins with a finite spontaneous curvature exhibit
a separated/corrugated (SC) phase, where hexagonal microdomains are formed,
in addition to the unbound and bound phases (see Fig.~\ref{fig:psep})~\cite{gout21}.
To lower the bending energy of the bound domains with $\kappa_{\rm p}>\kappa_{\rm d}$,
the unbound membrane bends unfavorably.
The changes from the unbound to SC phases and from the SC to bound phases are second-order and first-order transitions, respectively.
The interactions between proteins are not pairwise additive.
Since the membrane is largely curved, the interactions between the proteins are not expressed by those in a flat membrane;
therefore, membrane deformation must be explicitly included in theoretical analysis.
The SC phase can be analytically predicted by approximating it as a periodically curved one-dimensional stripe~\cite{gout21}.
Thus, the assumption of curved membrane shapes with a few parameters is useful for the analysis of microdomain structures.

For vesicles and membrane tubes, proteins with $C_0=0$
can induce microdomain structures~\cite{nogu16a} (see Figs.~\ref{fig:iso}(a)--(c)).
The bound membranes form flat domains, and instead, the unbound membranes are largely bent.
As a result, polyhedral vesicles and polygonal tubes are obtained.
Note that a tube-length constraint is required for the formation of polygonal tubes.
Under a constant axial force, these azimuthal phase separations are unstable;
instead, a first-order transition occurs between a wide tube with a high protein density and narrow tube with a low density at $f_{\rm ex}\sigma/k_{\rm B}T\simeq 20$ for  $\mu/k_{\rm B}T=3.5$.
However, for a fixed tube length, phase separation can also occur in the axial direction
 between an unbound membrane tube and a disk-like or polygonal tube, 
in which the total membrane area is too large and small, respectively (see  Fig.~\ref{fig:iso}(c)).
In general, 
 phase domains can be more easily formed in an ensemble of a constant extensive variable (here, tube length and number of proteins)
than in the ensemble in which the conjugated intensive variable is fixed, owing to a macroscopic phase separation (e.g., the vapor--liquid coexistence in the $NVT$ ensemble~\cite{wata12}).

For proteins with $C_0> 0$,
the microphase separation along the tube axis is obtained in addition to the bound and unbound phases,
even when $f_{\rm ex}$ and $\mu$ are fixed~\cite{nogu21b}
 (see Fig.~\ref{fig:iso}(d)).
When cylindrical tubes of the bound membranes are destabilized,
they deform into a bead-like round shape, and short cylindrical tubes of the unbound membranes are formed between them.
When the tube length is fixed, this phase separation occurs under a wider range of conditions.
At a small force $f_{\rm ex}$, the bound membranes in the tube necks are in contact, leading to vesicle formation.
In particular, at $f_{\rm ex}=0$, this deformation occurs through unduloid shapes~\cite{kenm03,nait95},
 which maintain a constant value of the mean curvature $H$ everywhere.
Similarly, for a vesicles, protein binding at $C_0> 0$ can induce vesicle division via budding.

Similar shapes of lipid domains have been observed in the experiments on protein-free membranes, 
where lipids are separated into disordered and ordered phases~\cite{baum03,veat03,yana08,yana10,chri09}.
In these systems, the difference in bending rigidity and the line tension of domain boundary 
are the main factors determining the domain shapes.
In contrast, differences in the spontaneous curvature play a crucial role for binding of curvature-inducing proteins.

\begin{figure*}[tb]
\includegraphics[width=16cm]{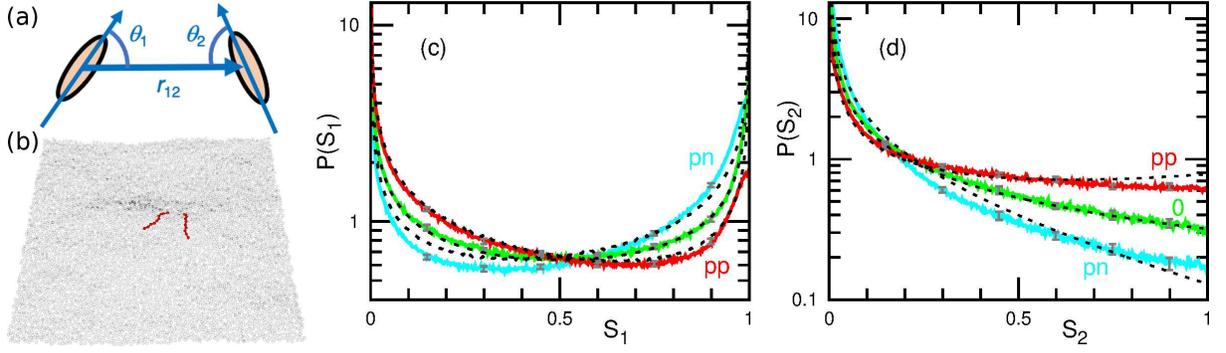}
\caption{
(Color online)
Interactions between two anisotropic proteins in a tensionless membrane ($\gamma=0$).
(a) Geometrical parameters for two proteins.
(b) Snapshot of the meshless membrane simulation. A protein rod consists of ten red particles.
(c),(d) Probability distribution of the orientational parameters 
(c) $S_1=\cos^2(\theta_1)$ and (d) $S_2=\sin^2(\theta_1)\sin^2(\theta_2)$.
Two protein rods are separated by the rod length, i.e., $r_{12}=r_{\rm {rod}}$.
The colored solid lines represent the simulation data 
for identical positive rod curvature $C_{\rm {r1}}=C_{\rm {r2}}=4/r_{\rm {rod}}$ (red, labeled by pp),
opposite curvatures $C_{\rm {r1}}= -C_{\rm {r2}}=4/r_{\rm {rod}}$ (cyan, labeled by pn), and zero curvature $C_{\rm {r1}}= C_{\rm {r2}}=0$ (green, labeled by 0).
The dashed lines represent the theoretical results.
Reproduced from Ref.~\citenum{nogu17} with permission from the Royal Society of Chemistry.
}
\label{fig:pair}
\end{figure*}

\subsection{Anisotropic proteins}\label{sec:asani}

\subsubsection{Interactions between two proteins}\label{sec:pair}

In this subsection, we consider the assembly of anisotropic proteins in membranes.
First, we describe the membrane-mediated interactions between two anisotropic proteins~\cite{nogu17}.
To simplify the theoretical calculations,  the proteins are treated as point-like objects~\cite{nogu17,domm99,domm02,yolc14}.
In a tensionless membrane ($\gamma=0$),
 the curvature-mediated interaction energy $H_{\rm int}(r_{12})=H_{\rm int}^{(0)}(r_{12})$ between two protein rods
 is given in the leading order by~\cite{nogu17}
\begin{eqnarray}
H_{\rm int}^{(0)}(r_{12}) &=&
\frac{16 \pi {r_{\rm rod}}^4}{9{r_{12}}^2}\kappa_{\rm d} C_{\rm r1} C_{\rm r2} \\ \nonumber
&&\times \big[\cos(2\theta_1)+\cos(2\theta_2)
 -\cos(2\theta_1-2\theta_2)\big]. \label{eq:Hint0}
\end{eqnarray}
Two proteins are rigid and have curvatures $C_{\rm r1}$ and $C_{\rm r2}$ with a length of $r_{\rm rod}$.
The angles $\theta_1$ and $\theta_2$  are shown in Fig.~\ref{fig:pair}.
This interaction is of a longer range than the Casimir-like interaction ($\propto\!{r_{12}}^{-4}$) between straight rods, 
which differently depends on the angles~\cite{gole96,bitb11}.

For $\theta_1=0$, the interaction energy (Eq.~(\ref{eq:Hint0})) is independent of $\theta_2$, as 
$H_{\rm int}^{(0)}(r_{12})= (16\pi {r_{\rm rod}}^4/9{r_{12}}^2)\kappa_{\rm d} C_{\rm r1}C_{\rm r2}$.
On the contrary, for $\theta_1=\theta_2=\pi/2$,
the energy has the opposite sign and the amplitude is three times larger as
$H_{\rm int}^{(0)}(r_{12})= -(16\pi {r_{\rm rod}}^4/3{r_{12}}^2)\kappa_{\rm d} C_{\rm r1}C_{\rm r2}$.
Hence, when two rods are identical, i.e., $C_{\rm r1}=C_{\rm r2}$,
they have a strong attractive interaction at $\theta_1=\theta_2=\pi/2$
and a weak repulsive interaction at $\theta_1=0$ or $\theta_2=0$.
When two identical rods contact side-by-side and form a dimer, the membrane undergoes less deformation, 
which generates this attraction.
For $C_{\rm r1}C_{\rm r2}<0$, the interactions are opposite.
For $\theta_1=0$ or $\theta_2=0$, the rods have a weak attractive interaction. 
Hence, rods with opposite curvatures prefer to be in a tip-to-tip contact.

For a positive surface tension $\gamma>0$, 
the effects of tension are dominant on a length scale larger than $\ell_{\rm t}=\sqrt{\kappa_{\rm d}/\gamma}$,
while
they are negligible on a length-scale smaller than $\ell_{\rm t}$.
The interaction energy exhibits a crossover from a bending-dominant regime to a tension-dominant regime at $r_{12}\approx3\ell_{\rm t}$~\cite{nogu17}: 
\begin{eqnarray}
H_{\rm int}(r_{12})=
&H_{\rm int}^{(0)}(r_{12}),& {\rm for~}r_{\rm rod}<r_{12}\ll \ell_{\rm t},
\\ \nonumber
&H_{\rm int}^{(1)}(r_{12}),& {\rm for~}r_{12}\gg\ell_{\rm t}{\rm ~if~}\cos[2(\theta_1-\theta_2)]\ne0,
\\ \nonumber
&H_{\rm int}^{(2)}(r_{12}),& {\rm for~}r_{12}\gg\ell_{\rm t}{\rm ~if~} \cos[2(\theta_1-\theta_2)]=0,
\end{eqnarray}
where
\begin{eqnarray}
H_{\rm int}^{(1)}(r_{12})&=&-\frac{64\pi {r_{\rm rod}}^4{\ell_{\rm t}}^2}{3{r_{12}}^4}\kappa_{\rm d}C_{\rm r1}C_{\rm r2}
\cos[2(\theta_1-\theta_2)],
\\ 
H_{\rm int}^{(2)}(r_{12})&=&
\frac{(2\pi)^{3/2}{r_{\rm rod}}^4}{9{\ell_{\rm t}}^{3/2}{r_{12}}^{1/2}}\exp\big(-\frac{r_{12}}{\ell_{\rm t}}\big)  \kappa_{\rm d} C_{\rm r1}C_{\rm r2} \big[2 \\ \nonumber
&& +2\cos(2\theta_1)+2\cos(2\theta_2)+\cos(2\theta_1+2\theta_2)
\big].
\end{eqnarray}
At the crossover, the energy changes from a $\sim\!{r_{12}}^{-2}$ power-law to a $\sim\!{r_{12}}^{-4}$ power-law (an exponential decay for $\theta_1-\theta_2 = (1/2 + n)\pi/2$).

These analytical results show good agreement with the simulation results, as shown in Fig.~\ref{fig:pair}~\cite{nogu17}.
The energy amplitude is scaled by a factor $\simeq\!1/20$, since the protein rods are flexible in the simulation
and the assumption of the rigid object in the theory gives an overestimation of the energy.
The angular distributions for an isolated pair of rods separated by a short distance $r_{12}=r_{\rm rod}$ are calculated as
 $P(\theta_1,\theta_2)\propto\exp(-H_{\rm int}/k_{\rm B}T)$.
Similar angular-dependent interactions have been reported for elliptic~\cite{schw15} and circular objects~\cite{kohy19a}.

Note that the protein axis is the direction of spontaneous curvature in this analysis,
which is not always the geometrical axis of the protein.
When the side part of a protein or inclusion strongly binds 
the membrane and the protein sinks in the membrane,
the protein instead bends the membrane in the side direction (i.e., $\kappa_{\rm side}{C_{\rm pside}}^2\gg \kappa_{\rm pm}{C_{\rm p}}^2$ in Eq.~(\ref{eq:ubrod})).
This is the mechanism of the tip-to-tip assembly reported in some coarse-grained molecular simulations~\cite{simu13,olin16} as discussed in Ref.~\citenum{nogu17}.
In the modeling procedure of coarse-grained proteins, side bending energy can be generated,
since a coarse-grained molecule typically has a more rounded shape than the original atomistic model.

\begin{figure*}[tb]
\includegraphics{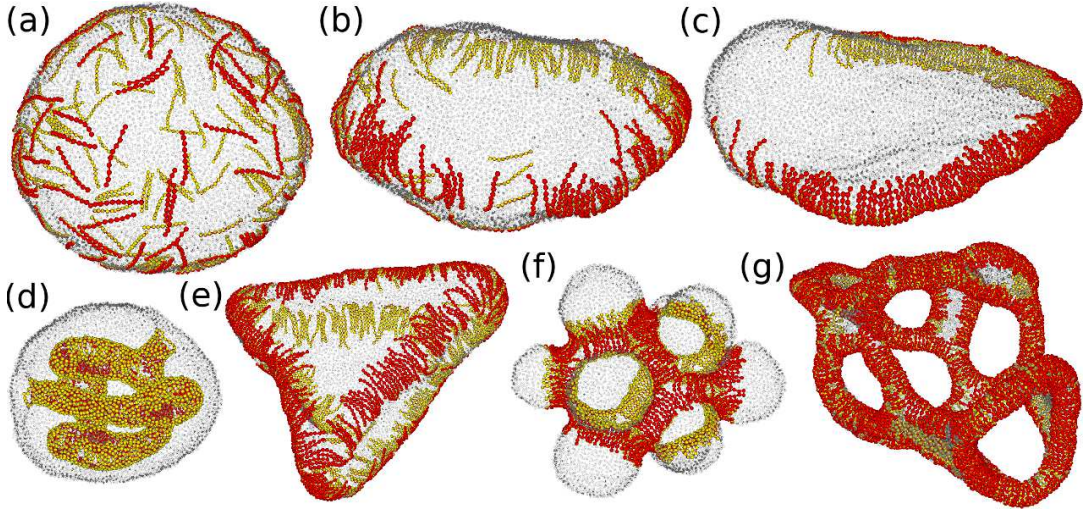}
\caption{
(Color online)
Snapshots of vesicles bound by protein rods with $r_{\rm rod}=10\sigma$.
(a)--(c) Protein assembly with increasing rod curvature $C_{\rm rod}$ at $\phi=0.167$~\cite{nogu15b,nogu16a,nogu14}:
(a) $C_{\rm rod}r_{\rm rod}=1.5$, (b) $C_{\rm rod}r_{\rm rod}=2.5$, and (c) $C_{\rm rod}r_{\rm rod}=3.5$.
(d) Tubular invagination for $C_{\rm rod}r_{\rm rod}=-4$ and $\phi=0.3$~\cite{nogu16}.
(e) Triangular-prism-shaped vesicle for $C_{\rm rod}r_{\rm rod}=2.5$ and $\phi=0.4$~\cite{nogu15b}.
(f) Multi-spindle-shaped vesicle for $C_{\rm rod}r_{\rm rod}=-4$ and side curvature $C_{\rm side}r_{\rm rod}=-1$ at $\phi=0.3$~\cite{nogu16}.
($C_{\rm side}=0$ is used for all other simulation results in the figures).
(g) High-genus vesicle formed by membrane rupture for  $C_{\rm rod}r_{\rm rod}=4$ and $\phi=0.8$~\cite{nogu16a}.
}
\label{fig:ves}
\end{figure*}

\subsubsection{Assembly in vesicles and tubes}\label{sec:rodves}

\begin{figure*}[tb]
\includegraphics{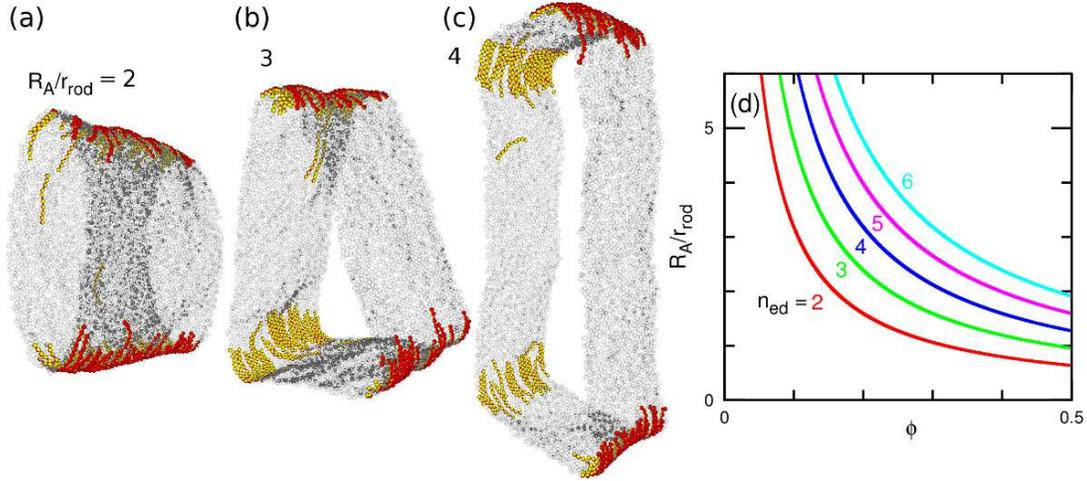}
\caption{
(Color online)
Formation of polygonal tubes by the protein rod assembly~\cite{nogu15b,nogu22}.
(a)--(c) Snapshots of (a) elliptic, (b) triangular, and (c) quadrangular tubes 
for $R_{\rm A}/r_{\rm rod}=2$, $3$, and $4$, respectively at $\phi=0.167$, $C_{\rm rod}r_{\rm rod}=2.5$, and $r_{\rm rod}=10\sigma$.
(d) Tube contour radius $R_{\rm A}$, where the edges of the polygonal tube with $n_{\rm ed}$ edges are filled with proteins,
given by Eq.~(\ref{eq:polygon}).
}
\label{fig:polycyl}
\end{figure*}

\begin{figure*}[tb]
\includegraphics[width=17cm]{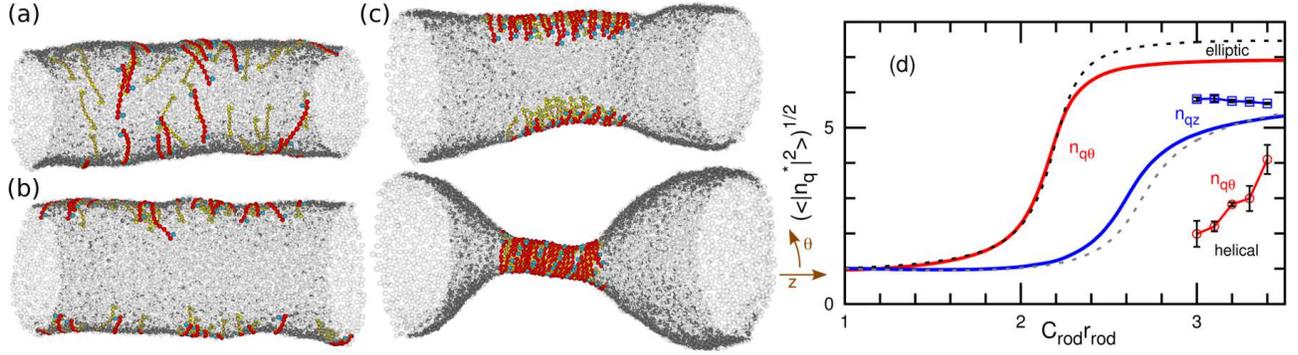}
\caption{
(Color online)
Membrane tube deformation induced by chiral and achiral protein rods  at $R_{\rm cyl}/r_{\rm rod}=1.31$ and $r_{\rm rod}=10\sigma$. 
(a)--(c) Snapshots at (a) $C_{\rm {rod}}r_{\rm {rod}}=1.5$, (b) $2.3$, 
and (c) $3.3$ for the chiral rods.
Two states coexist in (c) (an elliptic tube with two protein assemblies at the edges 
and a cylindrical tube with a helical protein assembly).
The membrane particles are displayed as transparent gray spheres.
The chiral or achiral protein rod has two additional (cyan) particles with the opposite or same side, respectively.
(d) Fourier amplitudes of protein densities.
The solid and dashed lines represent the data for the chiral and achiral protein rods, respectively.
The circles and squares with solid lines represent the azimuthal ($q\theta$) and axial ($qz$) modes, 
for the helical protein-assembly, respectively.
The Fourier amplitudes are normalized by the values at $C_{\rm {rod}}=0$. 
Reproduced from Ref.~\citenum{nogu19a}.
}
\label{fig:chcyl}
\end{figure*}

Protein assembly induces various shapes of vesicles and membrane tubes as shown in Figs.~\ref{fig:ves}--\ref{fig:chcyl}.
At a low protein density, proteins assemble with increasing rod curvature $C_{\rm rod}$ 
or bending rigidity $\kappa_{\rm p}$~\cite{nogu15b,nogu16a,nogu14} (see Figs.~\ref{fig:ves}(a)--(c)).
First, the vesicle deforms into an oblate shape, and the proteins are concentrated in the oblate equator; 
subsequently, the proteins assemble into a one-dimensional arc.
Hence, phase separation occurs as two continuous transitions in the axial and side directions of proteins
owing to the anisotropy of the bending energy.
With increasing protein density $\phi$,
the vesicle forms an elliptic disk, whose edges are covered by proteins.
With a further increase,
polyhedral vesicles are formed, and their highly curved edges are stabilized by the protein assembly~\cite{nogu15b} (see Fig.~\ref{fig:ves}(e)).

A protein with a negative curvature ($C_{\rm rod}<0$) can induce invaginations inside of the vesicle~\cite{nogu16} (see Fig.~\ref{fig:ves}(d));
these invaginations have tubular and disk-shaped regions like in the inner membrane of mitochondria~\cite{sche08,mann06}.
Similar tubular invagination was observed in experiments of BAR proteins.\cite{matt07}
Proteins with a negative side curvature prefer a saddle shape
and form a branched network in vesicles as well as in flat membranes~\cite{nogu16} (see Fig.~\ref{fig:ves}(f)).
The binding of proteins with high bending rigidity can induce membrane rupture, 
leading to the formation of high-genus vesicles~\cite{nogu16a} (see Fig.~\ref{fig:ves}(g)).
Similar high-genus vesicles have been observed in experiments and other simulations~\cite{ayto09}.

For membrane tubes, polygonal tubes are formed instead of polyhedrons (see Fig.~\ref{fig:polycyl}).
The proteins are concentrated on the edges of the polygon and are oriented in the  azimuthal direction,
 stabilizing the higher curvature of the edges and ensuring that the unbound membrane regions have lower curvatures.
Thus, the polygonal tubes can be spontaneously formed.
The geometrical condition required to fill the edges with proteins is given by
\begin{equation} \label{eq:polygon}
R_{\rm A}= \frac{n_{\rm ed}r_{\rm rod}}{2\pi\phi},
\end{equation}
where the contour radius $R_{\rm A}=A/2\pi L_z$, and $n_{\rm ed}$ is the number of edges.
This condition provides a rough estimate of the tube shape (see Fig.~\ref{fig:polycyl}).
When the proteins are slightly overfilled, the polygonal tubes are buckled~\cite{nogu15b}.
The phase boundary of the shapes can be more accurately estimated by taking the bending energy and mixing entropy
into account.
Note that  triangular prismatic tubes are observed in the inner membranes of mitochondria 
in astrocytes~\cite{sche08,blin65,fern83}.
The formation mechanism of these triangular tubes is not known, but
they may be generated by the assembly of BAR proteins.

\begin{figure*}[tb]
\includegraphics[width=17cm]{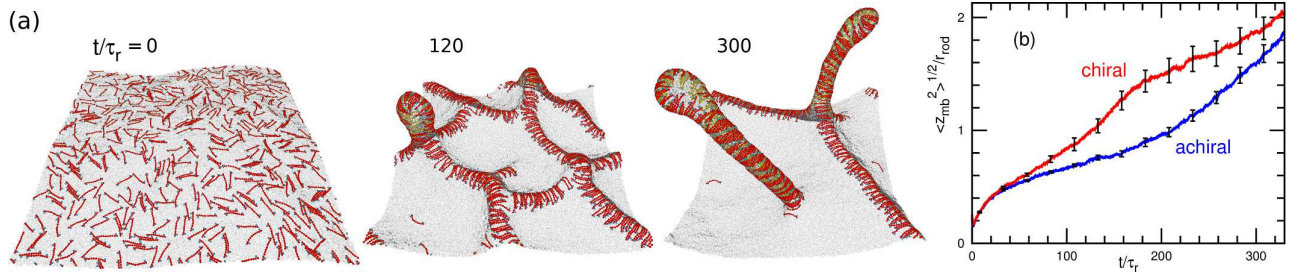}
\caption{
(Color online)
Tubulation from a flat tensionless membrane ($\gamma=0$) at $\phi_{\rm {rod}}=0.2$, $C_{\rm {rod}}r_{\rm {rod}}=2.5$, and $r_{\rm rod}=10\sigma$.
(a) Sequential snapshots for the chiral protein rods
at $t/\tau_{\rm r}= 0, 120$, and $300$, where $\tau_{\rm r}$ is the simulation time unit.
(b) Time development of the vertical membrane span $\langle {z_{\rm {mb}}}^2 \rangle^{1/2}$.
The red and blue lines show the data for the chiral and achiral rods, respectively.
Reproduced from Ref.~\citenum{nogu19a}.
}
\label{fig:chtube}
\end{figure*}

For a low density $\phi$,
the protein orientation changes from the tube-axial direction to the azimuthal direction with increasing protein curvature $C_{\rm rod}$~\cite{nogu15b,nogu16a,nogu14} (see Figs.~\ref{fig:n80}(a),(b)). The force $f_{\rm ex}$ along the tube is almost constant during this orientational change,
and this dependence can be well expressed by the theory described in Sec.~\ref{sec:tani}~\cite{nogu22}.
A further increase in  $C_{\rm rod}$ induces the two-step phase separation, as in the vesicle (see Fig.~\ref{fig:chcyl}); the tube deforms into an elliptic shape and the proteins are concentrated at the two edges. Subsequently, the proteins assemble in the tube axial direction, and the remainder of the tube returns to a circular shape~\cite{nogu15b,nogu16a,nogu14}.

In the simulations using a dynamically triangulated membrane model~\cite{rama12,rama13,rama18},
an anisotropic protein has been modeled as a single vertex with an orientational vector.
The proteins attractively interact with each other as a function of the distance between them 
and the angle between protein orientations.
Since a tip-to-tip pair of proteins has the same attraction as a side-by-side pair,
the proteins also assemble in the tip-to-tip direction.
Thus, a few layers of proteins form an arc-shaped edge in a vesicle
 in their simulations~\cite{rama12,rama13,rama18}, unlike the single layer seen in Fig.~\ref{fig:ves}(c), 
although their entire arc shapes are similar.
Thus, protein assembly and membrane shapes can be modified by the simulation models and potentials.

\begin{figure*}[tb]
\includegraphics[width=16cm]{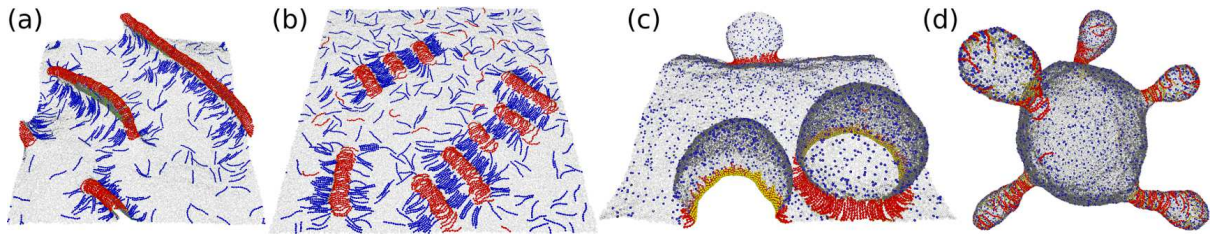}
\caption{
(Color online)
Snapshots of membranes with two types of proteins~\cite{nogu17,nogu17a}.
(a),(b) Mixture of two types of protein rods with positive rod curvature $C_{\rm rod}r_{\rm rod}=4$ 
and negative rod curvature $C_{\rm rod}r_{\rm rod}=-2$ 
for $\phi_{\rm pos}=0.1$, $\phi_{\rm neg}=0.2$, and $r_{\rm rod}=10\sigma$
 at (a) $\gamma=0$ and (b) $\gamma {r_{\rm rod}}^2/k_{\rm B}T=100$.
(c),(d) Mixture of the protein rods of $r_{\rm rod}=10\sigma$ and 
 isotropic proteins at  $\phi_{\rm rod}=0.1$ and $\phi_{\rm iso}=0.1$.
(c) Tensionless membrane ($\gamma=0$) for $C_{\rm rod}r_{\rm rod}=-3$ and  $C_{\rm iso}r_{\rm rod}=2$. 
(d) Vesicle for  $C_{\rm rod}r_{\rm rod}=3$ and $C_{\rm iso}r_{\rm rod}=3$. 
A protein rod is displayed as
a chain of spheres whose halves are colored
in red and yellow (or in blue and green).
The isotropic proteins and unbound membrane are  displayed 
in blue and transparent gray, respectively.
}
\label{fig:mix}
\end{figure*}

\subsubsection{Chirality of proteins}\label{sec:chiral}

BAR domains have chirality, and their helical alignments on the membrane tubes have been observed by electron microscopy~\cite{mim12a,fros08,adam15}.
The pitch and width of the alignment vary according to the type of BAR domain.
We investigated the chirality effects by modifying the protein model in our meshless membrane simulation~\cite{nogu19a}.
Two beads connected to the protein tips are added on the opposite or same side
for the chiral or achiral protein model, respectively.
The chirality does not affect the protein behavior under isolated conditions
or in the protein assembly in the edges of the elliptic membrane (compare solid and dashed lines 
in Fig.~\ref{fig:chcyl}(d)).
However, chirality additionally induces a first-order transition for high rod curvature $C_{\rm {rod}}$
(see the two membrane shapes in Fig.~\ref{fig:chcyl}(c) 
and the lines with symbols in Fig.~\ref{fig:chcyl}(d)).
Two protein assemblies of the elliptic edges are unified into a helical alignment, and the membrane forms a cylindrical tube.
The tube radius of this assembly is determined by the rod curvature, in agreement with the experimental evidence that each type of BAR protein typically generates a constant radius of the membrane tubules~\cite{itoh06,masu10,mim12a}.
The side-by-side attractive interaction between the protein rods increases the stability of this assembly~\cite{nogu19a}.

Membrane tubules can be spontaneously formed by the protein assembly from a flat membrane~\cite{nogu16,nogu17a} (see Fig.~\ref{fig:chtube}).
The chirality promotes this tubulation~\cite{nogu19a} (see Fig.~\ref{fig:chtube}(b));
in the helical alignment, the proteins can interact with other proteins in both the tip and side directions,
so their assembly efficiently generates tubules.
Moreover, the tubulation is suppressed by the positive surface tension ($\gamma>0$) and percolated network formation due to the negative side curvature of the protein rods~\cite{nogu16}.

\subsection{Mixture of multiple types of proteins}\label{sec:mix}

Up to this point, we have described the assembly of a single type of protein.
However, many types of proteins work cooperatively to bend biomembranes in living cells.
For instance, in clathrin-mediated endocytosis, BAR proteins bind the membrane, and subsequently, the clathrin-coat forms a spherical bud.
Later, the bud is pinched off by the binding of dynamin to the bud neck~\cite{mcma11,kaks18,schm11,anto16}.
In this last subsection,
we consider the mixture of different types of proteins.

Two types of protein rods with opposite rod curvatures are repulsive in the side-by-side direction
but weakly attractive in the tip-to-tip direction, as described in Sec.~\ref{sec:pair}.
This attractive interaction becomes large between the protein assemblies (see Figs.~\ref{fig:mix}(a),(b)).
Each type of protein forms a one-dimensional assembly in which the neighboring proteins have the side-by-side contact.
The protein assembly of the other type (opposite rod curvature) is attached to the lateral sides of this assembly
and stabilizes straight bump  and stripe structures~\cite{nogu17}.
Tubulation is prevented by this bump-shaped assembly.
When one type of protein has a low-amplitude rod curvature (blue rods in Fig.~\ref{fig:mix}(a)),
the bump exists in an isolated manner at low surface tension.
On the contrary, at high surface tension or high-amplitude rod curvature, the bumps attract each other, leading to a periodic bump structure (see Fig.~\ref{fig:mix}(b)).
Since a flatter structure has a larger projected area,
 flatter structures, such as this stripe and hexagonal network structures, are generally formed under higher surface tensions~\cite{nogu17,nogu16}.

When protein rods and isotropic proteins, represented by one membrane particle with high bending rigidity and nonzero spontaneous curvature, are mixed, spherical buds are formed~\cite{nogu17a} (see Figs.~\ref{fig:mix}(c),(d)).
Protein rods assemble at the necks of the buds, either aligning perpendicularly (Fig.~\ref{fig:mix}(c)) 
or parallel  (Fig.~\ref{fig:mix}(d)) to the azimuthal direction.
The former alignment was considered in an experiment on Pacsin~\cite{shim10}.
BAR proteins or polymerized dynamin surround
the neck of an endocytotic bud by the latter assembly~\cite{suet14,mcma11}.

Tubulation can be promoted or suppressed by the addition of isotropic proteins with the same or opposite sign of the spontaneous curvature, respectively~\cite{nogu17a}.
When the same amounts of two types of isotropic proteins with opposite spontaneous curvatures are added,
tubulation is promoted in the late stage, since the isotropic proteins become locally concentrated owing to the curvature sensing.

\begin{figure*}[tb]
\includegraphics[width=16cm]{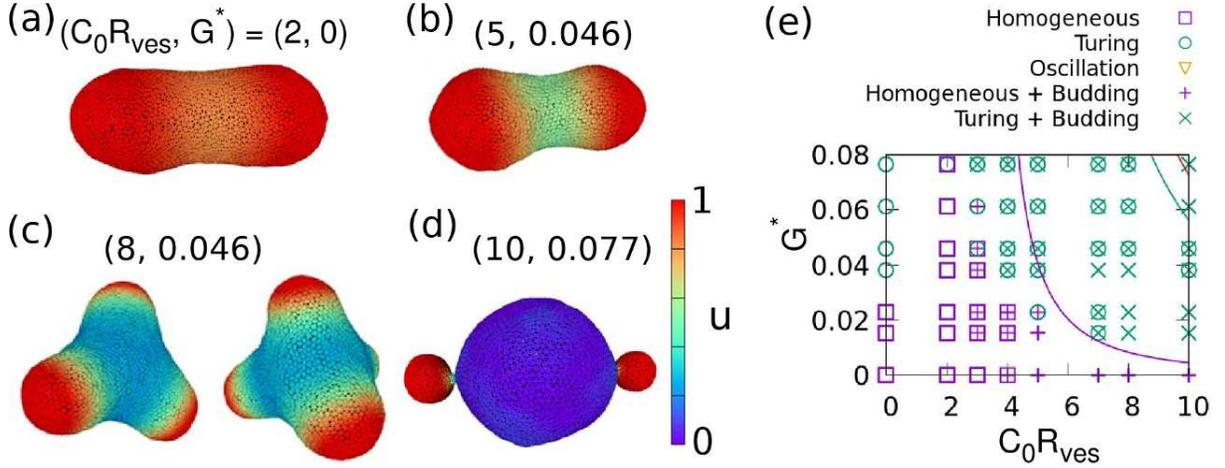}
\caption{
(Color online)
Pattern formation of a vesicle under coupling with a reaction-diffusion dynamics
at $V^*=0.8$ and $\kappa_{\rm p}/\kappa_{\rm d}=2$.
(a)--(d) Snapshots of vesicles at (a) $(C_0R_{\rm ves}, G^*)=(2,0)$, (b) $(5,0.046)$, (c) $(8,0.046)$, and (d) $(10,0.077)$,
where $G^*= G\kappa_{\rm d}/{R_{\rm ves}}^2$.
(c) Three- and four-spindle shapes are formed from initially prolate and oblate vesicles, respectively.
The color indicates the concentration $u$ of the curvature-inducing proteins.
(e) Phase diagram, where
the overlapped symbols indicate the coexistence of multiple patterns.
The purple and green curves represent the Turing and Hopf bifurcations, respectively,
which are analytically determined for a non-deformable spherical vesicle with a radius of $R_{\rm ves}$.
Reproduced from Ref.~\citenum{tame20}.
}
\label{fig:turing}
\end{figure*}

\section{Coupling with Reaction-Diffusion Dynamics}\label{sec:rd}

We have reviewed protein binding and membrane deformations in
thermal equilibrium or metastable states and relaxation into them until here.
In living cells, biomembranes are usually under nonequilibrium conditions,
and their shapes are often kinetically determined.
Membrane undulations are modified  under nonequilibrium~\cite{pros96,mann01,turl16,alme17,nogu21}.
Red blood cells and liposomes with F$_1$F$_0$-ATPase exhibit larger fluctuations~\cite{turl16,alme17},
whereas a moving membrane pushed by protein filament growth can also suppress fluctuations~\cite{nogu21}.
Protein binding and unbinding rates can deviate from a detailed balance~\cite{gout21}.
Moreover, protein (un)binding is activated or inhibited by other proteins in living cells.
Such dynamics can be described by reaction-diffusion equations.
Membrane deformations accompanied by a wave of 
protein concentrations have been observed under various conditions~\cite{gov18,wu21,alla13,pele11,wu18,lits18,chri21}.
In this section, we consider the coupling of membrane deformation and reaction-diffusion system~\cite{tame20,tame21,tame22}.

\subsection{Coupling models}\label{sec:cmodel}

Two types of proteins are considered: curvature-inducing and regulatory proteins.
Their concentrations are $u$ and $v$, respectively.
Curvature-inducing proteins induce high bending rigidity and spontaneous curvature as described
by Eq.~(\ref{eq:Fcv0}) but neglecting the Gaussian curvature term (i.e., $\bar{\kappa}_{\rm p}=\bar{\kappa}_{\rm d}$).
Regulatory proteins do not directly modify the membrane properties.
The reaction-diffusion equations are given by
\begin{eqnarray}
\tau_{\rm rd}\frac{\partial u}{\partial t}&=& f(u,v) + D_{u} \nabla^2 u,  \\
\tau_{\rm rd}\frac{\partial v}{\partial t}&=& g(u,v) + D_{v} \nabla^2 v,
\end{eqnarray}
where $\tau_{\rm rd}$ is the reaction time unit and $D_{u}$ and $D_{v}$ are diffusion constants.

As reaction-diffusion models,
the Brusselator model~\cite{prig68} and the FitzHugh--Nagumo model~\cite{fitz61,nagu62}
are employed in Refs.~\citenum{tame20,tame21} 
and Ref.~\citenum{tame22}, respectively. 
In Ref.~\citenum{tame20},
the binding rate of $u$ is linearly dependent on the bending energy change as follows:
\begin{eqnarray}
f(u,v) &=& A- G\frac{\partial f_{\rm cv}}{\partial u} - (B+1)u + u^2v,  \label{eq:f1} \\ \label{eq:f2}
g(u,v) &=& Bu - u^2v,
\end{eqnarray}
where $f_{\rm cv}$ denotes the local bending energy per unit area.
As the membrane curvature approaches the preferred curvature,
the curvature-inducing protein ($u$) binds more frequently.
The regulatory protein ($v$) does not directly change the membrane properties.

In these studies~\cite{tame20,tame21,tame22},
a dynamically triangulated membrane model~\cite{nogu09,gomp04c,gomp97f,nogu05}
is combined with these reaction-diffusion models.
The membrane motion is solved by the Langevin equation~\cite{nogu05}
and the reaction-diffusion equations are solved by the finite-volume scheme~\cite{tame20}.

\subsection{Turing patterns}\label{sec:turing}

Figure~\ref{fig:turing} shows the phase diagram and typical vesicle shapes obtained in Ref.~\citenum{tame20}.
For  a non-deformable curved surface,
the boundaries of the Turing pattern and temporal oscillation mode are determined by linear stability analysis (see solid lines in Fig.~\ref{fig:turing}(e)).
However, these conditions are modified for deformable vesicles.
The Turing patterns are stabilized by membrane deformation, and
the region of the Turing patterns increases in the phase diagram.
Budding and multi-spindle shapes are also induced by Turing patterns (see Figs.~\ref{fig:turing}(c),(d)).
Hysteresis of vesicle shapes exists; initial oblate vesicles result in a larger number of spindles than prolate vesicles.
The number of spindles also increases with deceasing wavelength of the Turing patterns.
For budded vesicles,
a Turing domain boundary separating two phases with high and low values of $u$ is formed at the connective neck, 
because the diffusion of the proteins is reduced at the narrow neck.
Moreover, it is observed that a temporal oscillation of the protein concentration is changed into a Turing pattern by budding.

\begin{figure*}[tb]
\includegraphics[width=16cm]{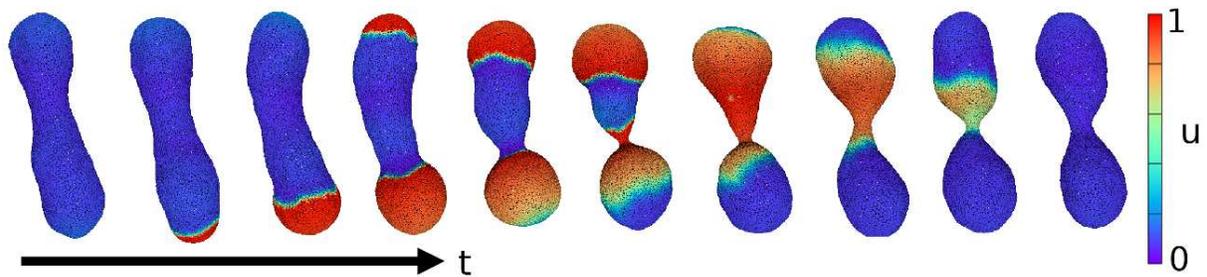}
\caption{
(Color online)
Sequential snapshots of a vesicle under coupling with a reaction-diffusion dynamics
under a slow reaction condition ($\tau_{\rm rd}=400\tau_{\rm md}$)
at $V^*=0.65$, $\kappa_{\rm p}/\kappa_{\rm d}=4$, and $C_0R_{\rm ves}=4$.
The color indicates the concentration $u$ of the curvature-inducing proteins.
Each frame step is $\tau_{\rm rd}$.
The chemical wave is accompanied by the vesicle shape oscillation.
Reproduced from Ref.~\citenum{tame21} with permission from the Royal Society of Chemistry.
}
\label{fig:wave}
\end{figure*}

\subsection{Traveling waves}\label{sec:wave}

The vesicle shape can spontaneously oscillate, associated with the reaction-diffusion waves of curvature-inducing proteins as shown in Fig.~\ref{fig:wave}~\cite{tame21}.
Similar shape oscillations have been  experimentally observed for liposomes with a reconstituted Min system~\cite{lits18,chri21}. 
Min proteins exhibit a traveling wave on membranes, which is considered to play an important role in cell division in Escherichia coli~\cite{ramm19,lutk12}.

Moreover,  traveling waves can spontaneously be triggered at a high- or low-curved membrane region
depending on the coupling parameters, even under the condition of a spatially homogeneous oscillation on a sphere~\cite{tame21}.
On tubular membranes, excitable waves can deform the membrane into meandering shapes~\cite{tame22};
the wave speed depends on the membrane curvature, and the necked shape of the membrane can erase the wave.
 
We have explained several membrane dynamics coupled with reaction-diffusion systems, 
using two types of well-known reaction models.
However, there are large degrees of freedom in reaction and coupling models. 
Qualitatively different membrane dynamics may occur when other models are applied.
For example, when a curvature-inducing protein is used as an inhibitor,
the dynamics may change significantly. 
Further studies are required to better understand such systems.

\section{Outlook}\label{sec:out}

We have reviewed the interactions of membranes with curvature-inducing proteins.
These proteins induce many types of membrane deformation.
Understanding of these systems has progressed intensively in recent studies.
To conclude this paper, we provide remarks regarding further studies.

\noindent
(i) Mechanical properties of proteins\\
The bending rigidity of bound proteins is an important quantity, as described in Secs.~\ref{sec:tiso} and \ref{sec:tani}.
Nevertheless, they have not been estimated for many proteins,
and completely rigid protein models have often been used in theories and simulations.
Curvature generation can be overestimated by using this rigid-body approximation.
Thus, the estimation of the bending rigidity by experiments and molecular simulations is important.
Although several research groups have reported all-atom and coarse-grained molecular simulations of BAR domains~\cite{bloo06,arkh08,yu13,take17,mahm19},
 bending rigidity has not been calculated.

Some anisotropic proteins may have a significant value of the side spontaneous curvature, which changes the protein-membrane interactions.
Recent experiments~\cite{zeno19,busc15} have shown that intrinsically unfolded domains of proteins enhance curvature sensing and induce the formation of small vesicles.
The interaction between such anchored chains and membrane generate an isotropic spontaneous curvature~\cite{bick06,hier96,auth03,auth05,evan03a,wern10} and
repulsion between them can stabilize microdomains due to the conformational entropy of chains~\cite{wu13}.
Tubulation and vesicle deformation are also significantly modified by the addition of anchored excluded-volume chains to chiral crescent protein rods~\cite{nogu22b}.

Here, we consider that bound membranes remain in a fluid phase.
However, the assemblies of clathrin, COPI, and COPII~\cite{fain13,beth18,kaks18,otte11} form regular lattice structures.
Moreover, a single clathrin triskelion does not sense membrane curvature but their assembly does.\cite{zeno21}
The solid structure may be needed to take into account to understand the budding induced by these proteins more quantitatively.

\noindent
(ii) Interactions with protein filaments and geometrical effects\\
In living cells, membranes often interact with
protein filaments (e.g., actin~\cite{svit18,skru20,suet12} and microtubules~\cite{wade09,dogt05}).
In particular, actin polymerization is involved in the formation of traveling waves~\cite{gov18,alla13,wu18,inag17,dene18}.
These filaments can push the membranes perpendicularly and also pull them laterally,
so that they can induce membrane protrusion and oppositely flatten the membrane depending on the manner of interaction.
Although the membrane--filament interactions have been investigated, many open questions remain.

Supported membranes, in which membranes are placed on a solid or polymer layer,  
have been experimentally used for a wide range of surface-specific analytical techniques~\cite{tana05,weer15}.
Recently, detachment of the lipid membrane from the substrate induced by the binding of annexins has been reported~\cite{boye17,boye18}.
Various dynamics, such as rolling and budding, have been observed.
Although budding and subsequent vesicle formation were simulated by the meshless simulation of the membrane with a homogeneous spontaneous curvature~\cite{nogu19c},
 rolling has not yet been reproduced.
Thus, the interactions between curvature-inducing proteins and membranes with open edges should be further explored.

Organelle shapes are regulated by many types of curvature-inducing proteins and filaments.
In addition, volume control by osmotic pressure and geometrical constraints plays a role in determining their shapes.
Mitochondria consist of two bilayer membranes,
and the inner membrane has a much larger surface area than the outer membrane~\cite{sche08,mann06}.
When a vesicle exists inside another vesicle with a smaller surface area,
this geometrical constraint can induce an invagination similar to that in the mitochondria~\cite{kahr12a,kahr12b,saka14,kavc19}.
The nuclear envelope shape~\cite{gros12,ungr17} can be understood as the stomatocyte of a high-genus vesicle with pore size constraint by the nuclear pore complex~\cite{nogu16b}.
The membrane deformation by curvature-inducing proteins under geometrical constraints is an important problem for further studies.

\noindent
(iii) Cooperation of multiple types of proteins\\
We described the interactions of two or three types of curvature-inducing proteins
and the reaction-diffusion systems of two types of proteins  in Sec.~\ref{sec:mix} and in Sec.~\ref{sec:rd}, respectively.
In living cells, many more types of proteins are involved in shape regulation.
During endocytosis, many types of proteins work cooperatively~\cite{suet14,mcma11,kaks18,schm11,raib09,daum14}.
Recently, it was reported that the curvature sensing of clathrin assembly  is enhanced in the presence of adaptor proteins.\cite{zeno21}
However, the cooperation  mechanism of different proteins in biological processes is much less understood 
 than that of each protein.
Their interactions under nonequilibrium conditions are important topics for further study.

\begin{acknowledgments}
This work was supported by JSPS KAKENHI Grant Number JP21K03481.
\end{acknowledgments}

\end{document}